\documentclass[11pt,a4paper]{article}

\usepackage[numbers, square, comma, sort&compress]{natbib}
\usepackage{amssymb}
\usepackage{amsthm}
\usepackage{amstext}
\usepackage{amsmath}
\usepackage{amsfonts}
\usepackage{empheq}
\usepackage{verbatim}
\usepackage{geometry}
\usepackage{latexsym}
\usepackage{t1enc}
\usepackage{graphicx}
\usepackage{color}
\usepackage{colortbl}
\usepackage[all]{xy}
\usepackage{makeidx}
\usepackage{slashed}
\usepackage{multicol}
\usepackage{alltt}

\allowdisplaybreaks

\renewcommand{\vec}[1]{\mbox{\boldmath$ #1 $}}

\newcommand{\be}{\begin{equation}}
\newcommand{\ee}{\end{equation}}
\newcommand{\bea}{\begin{eqnarray}}
\newcommand{\eea}{\end{eqnarray}}

\newcommand{\e}{\epsilon}
\newcommand \slsh [1] {\not\!{#1}}
\newcommand{\secn}[1]{Section~\ref{#1}}

\def\beq{\begin{equation}}
\def\eeq{\end{equation}}
\def\beqa{\begin{eqnarray}}
\def\eeqa{\end{eqnarray}}
\def\eq#1{Eq.~(\ref{#1})}

\def\spa#1.#2{\left\langle#1\,#2\right\rangle}
\def\spb#1.#2{\left[#1\,#2\right]}
\def\spash#1.#2{\spa{\smash{#1}}.{\smash{#2}}}
\def\spbsh#1.#2{\spb{\smash{#1}}.{\smash{#2}}}
\def\sand#1.#2.#3{%
  \left\langle\smash{#1^{-}}{\vphantom1}\right|{#2}%
  \left|\smash{#3^{-}}{\vphantom1}\right\rangle}
\def\sandp#1.#2.#3{%
  \left\langle\smash{#1^{-}}{\vphantom1}\right|{#2}%
  \left|\smash{#3^{+}}{\vphantom1}\right\rangle}
\def\sandpp#1.#2.#3{%
  \left\langle\smash{#1^{+}}{\vphantom1}\right|{#2}%
  \left|\smash{#3^{+}}{\vphantom1}\right\rangle}
\def\sandpm#1.#2.#3{%
  \left\langle\smash{#1^{+}}{\vphantom1}\right|{#2}%
  \left|\smash{#3^{-}}{\vphantom1}\right\rangle}
\def\sandmp#1.#2.#3{%
  \left\langle\smash{#1^{-}}{\vphantom1}\right|{#2}%
  \left|\smash{#3^{+}}{\vphantom1}\right\rangle}

\def\ssand#1.#2.#3{%
  \left\langle\smash{#1}{\vphantom1}\right|{#2}%
  \left|\smash{#3}{\vphantom1}\right]}
\def\ssandp#1.#2.#3{%
  \left\langle\smash{#1}{\vphantom1}\right|{#2}%
  \left|\smash{#3}{\vphantom1}\right\rangle}
\def\ssandpp#1.#2.#3{%
  \left\langle\smash{#1}{\vphantom1}\right|{#2}%
  \left|\smash{#3}{\vphantom1}\right\rangle}

\def\proj{\flat}
\def\projdot#1.#2{k_{#1}^\proj\cdot k_{#2}^\proj}
\def\sandff#1.#2.#3{%
  \left\langle\smash{#1^{\proj,-}}{\vphantom1}\right|{#2}%
  \left|\smash{#3^{\proj,-}}{\vphantom1}\right\rangle}
\def\sandnf#1.#2.#3{%
  \left\langle\smash{#1^{-}}{\vphantom1}\right|{#2}%
  \left|\smash{#3^{\proj,-}}{\vphantom1}\right\rangle}
\def\sandfn#1.#2.#3{%
  \left\langle\smash{#1^{\proj,-}}{\vphantom1}\right|{#2}%
  \left|\smash{#3^{-}}{\vphantom1}\right\rangle}

\def\spa#1.#2{\left\langle#1\,#2\right\rangle}
\def\spb#1.#2{\left[#1\,#2\right]}

\numberwithin{equation}{section}

\begin{document}


\begin{titlepage}

\hbox{NIKHEF/2017-25}
\hbox{Edinburgh 2017/10}
\hbox{QMUL-PH-17-07}
\hbox{ARC-17-03}

\vskip 2cm

\begin{center}
\Large{\sc{Universality of next-to-leading power threshold effects for colourless final states
in hadronic collisions}}
\end{center}

\vskip 8mm

\begin{center}

V. Del Duca$^{1,2}$, E.~Laenen$^{3,4,5}$, L.~Magnea$^6$, 
L.~Vernazza$^7$, C.~D.~White$^8$ \\ [6mm]

\vspace{6mm}

\textit{$^1$ETH Zurich, Institut fur theoretische Physik, Wolfgang-Paulistr. 27, 8093, 
Zurich, Switzerland}\\
\vspace{1mm}

\textit{$^2$INFN Laboratori Nazionali di Frascati, 00044 Frascati (Roma), Italy}\\
\vspace{1mm}

\textit{$^3$Nikhef, Science Park 105, NL--1098 XG Amsterdam, The Netherlands} \\ 
\vspace{1mm}

\textit{$^4$ITFA, University of Amsterdam, Science Park 904, Amsterdam, 
The Netherlands} \\
\vspace{1mm}

\textit{$^5$ITF, Utrecht University, Leuvenlaan 4, Utrecht, The Netherlands} \\
\vspace{1mm} 

\textit{$^6$Dipartimento di Fisica and Arnold-Regge Center, Universit\`a di Torino, \\
and INFN, Sezione di Torino, Via P. Giuria 1, I-10125 Torino, Italy} \\
\vspace{1mm}

\textit{$^7$Higgs Centre for Theoretical Physics, School of Physics and Astronomy, 
The University of Edinburgh, Edinburgh EH9 3JZ, Scotland, UK}\\
\vspace{1mm}

\textit{$^8$Centre for Research in String Theory, School of Physics and Astronomy, 
Queen Mary University of London, 327 Mile End Road, London E1 4NS, UK} \\
\vspace{1mm}

\end{center}

\vspace{5mm}

\begin{abstract}

\noindent
We consider the production of an arbitrary number of colour-singlet
particles near partonic threshold, and show that next-to-leading order
cross sections for this class of processes have a simple universal
form at next-to-leading power (NLP) in the energy of the emitted gluon
radiation. Our analysis relies on a recently derived factorisation
formula for NLP threshold effects at amplitude level, and therefore
applies both if the leading-order process is tree-level and if it is
loop-induced. It holds for differential distributions as well. The
results can furthermore be seen as applications of recently derived
next-to-soft theorems for gauge theory amplitudes. We use our
universal expression to re-derive known results for the production of
up to three Higgs bosons at NLO in the large top mass limit, and for
the hadro-production of a pair of electroweak gauge bosons. Finally,
we present new analytic results for Higgs boson pair production at NLO
and NLP, with exact top-mass dependence.

\end{abstract}

\end{titlepage}


\section{Introduction}
\label{sec:introduction}

The remarkable experimental precision and high statistics offered by current 
and future colliders necessitates the calculation of high-order effects in QCD 
perturbation theory for processes of increasing complexity. The calculation of 
next-to-leading order (NLO) QCD corrections has reached the stage where 
automated tools are being used and multi-jet production rates are evaluated 
(see, for example~\cite{Badger:2016bpw}). Attention then focuses on QCD 
corrections at NNLO or above, and on the inclusion of higher-order electroweak 
effects. In this context, processes which are induced by loop effects present 
special difficulties: typically, the leading order contribution involves a loop 
containing heavy particles, such as top quarks or electroweak vector bosons, 
and is often computed in the context of an effective field theory. NLO QCD 
corrections with exact dependence on the heavy particle masses then involve 
intricate two-loop calculations: for multi-particle final states, these corrections 
are not known, and even for two-particle final states they are typically only 
known approximately, or in some cases numerically. Clearly, in all these 
cases it is desirable, where possible, to improve upon existing results by 
providing analytic information. Even partial information can be useful, 
since it can be used, for example, to speed up numerical codes, and also 
to provide additional consistency checks. 

In this paper, we consider the production of a generic colour-singlet final state
in hadronic collisions, and we study the effects of additional gluon radiation 
near partonic threshold, where emitted gluons have a low energy or transverse 
momentum with respect to the incoming partons. In such cases, one commonly
defines a dimensionless {\it threshold variable} $\xi$, vanishing at the threshold, 
and it is well known that the corresponding differential cross section has the 
generic form
\beq
  \frac{d \sigma}{d \xi} \, = \, K_{\rm ew} \left( 4 \pi \alpha_s \right)^{n_0}
  \sum_{n = 0}^{\infty} \left( \frac{\alpha_s}{\pi} \right)^n 
  \sum_{m = 0}^{2 n - 1} \left[ \, c_{n m}^{(-1)} \left( \frac{\log^m \xi}{\xi} \right)_+ 
  + \, c_{n}^{(\delta)} \, \delta(\xi) + \, c_{nm}^{(0)} \, \log^m \xi + \ldots \, \right] ,
\label{thresholddef}
\eeq
where the overall factor is associated with the leading-order cross section: for 
loop-induced processes, this may be proportional to a power of the strong 
coupling, as indicated, while $K_{\rm ew}$ contains electroweak couplings.
The first two sets of terms on the right-hand side of \eq{thresholddef} originate 
from soft and collinear radiation (real or virtual), and correspond to the leading 
power in the threshold variable, and to corrections localised at the threshold, 
respectively. These contributions are known to have a universal (process-independent) 
form, that permits their resummation to all orders in perturbation theory. This 
resummation is well understood and widely applied, and can be performed 
within a variety of approaches (see for example \cite{Sterman:1986aj,Catani:1989ne,
Korchemsky:1993uz,Korchemsky:1993xv,Contopanagos:1997nh,Forte:2002ni,
Eynck:2003fn,Banfi:2004yd,Becher:2006nr,Luisoni:2015xha}). Even without a 
full resummation, a fixed order evaluation of these terms can be useful in 
estimating higher-order corrections to the cross-section, when these are 
not known (see for example~\cite{Kidonakis:2013dja} for a review). 

The third set of terms on the right-hand side of \eq{thresholddef}
defines {\it next-to-leading power} (NLP) contributions in the
threshold variable, roughly corresponding to gluon radiation that can
be {\it next-to-soft} or collinear. Although power-suppressed, these
terms are still singular as $\xi \rightarrow 0$, and can be
numerically significant: for example, they contribute significantly to
the theoretical uncertainty for Higgs boson production in the gluon
fusion channel~\cite{Kramer:1996iq}, and their numerical impact has
been explicitly confirmed by the recent calculation of this process at
N$^3$LO~\cite{Anastasiou:2013srw,Anastasiou:2013mca,
  Anastasiou:2014vaa,Anastasiou:2014lda,Anastasiou:2015ema}. A full,
generally applicable resummation prescription for NLP contributions is
not presently known, even in the relatively simple case of parton
annihilation into electroweak final states: the problem has been
intensively studied in recent years, and partial progress has been
made using a variety of methods~\cite{Laenen:2008ux,Laenen:2008gt,
  Laenen:2010uz,Bonocore:2014wua,Bonocore:2015esa,Bonocore:2016awd,Moch:2009mu,
  Moch:2009hr,Soar:2009yh,Almasy:2010wn,Presti:2014lqa,deFlorian:2014vta,
  Grunberg:2009yi,Grunberg:2009vs,Larkoski:2014bxa,Kolodrubetz:2016uim,
  Moult:2016fqy,Moult:2017rpl,Feige:2017zci,Gervais:2017yxv,Gervais:2017zky},
buiding upon the earlier work of~\cite{Low:1958sn,Burnett:1967km,
  DelDuca:1990gz}. Even without a full resummation, however, knowledge
of NLP contributions at fixed order can provide useful analytic
information where this is missing, as well as furnishing improved
approximations for unknown higher-order cross sections. This is
especially true if one can derive universal properties of NLP
contributions, applicable at a given order for a broad class of
processes.

In this paper, we examine the universal properties of NLP radiation in the hadro-production 
of an arbitrary number of colour-singlet particles at NLO. We will prove that the NLO 
cross-section for this class of processes can be written in terms of the leading-order 
cross-section with shifted kinematics, convolved with a simple universal $K$-factor. 
More precisely, we find that the $K$-factor does not depend on the spin of the emitting
parton, and depends on the color representation only through a trivial replacement 
of colour factors. Our starting point will be a factorisation formula for NLP effects at 
amplitude level recently derived in Refs.~\cite{Bonocore:2015esa,Bonocore:2016awd}, 
which expresses the effect of adding an additional gluon to an arbitrary hard process 
with an electroweak final state in terms of universal functions. Our results provide a 
useful testing ground for this formula, and an examination of its simpler consequences, 
aside from its role as a basis for resummation. Furthermore, our results have more 
practical consequences, providing an analytic approximation to the NLO cross section
for a number of interesting loop-induced processes for which only limited information
is available. Interestingly, the universality of the result extends to differential distributions,
provided the shift in LO kinematics is properly understood. From a theoretical point of 
view, the universality and simplicity of our results at NLO can be seen as a consequence
of recently derived next-to-soft theorems~\cite{Casali:2014xpa,White:2011yy,
Cachazo:2014fwa} for radiative tree-level gauge theory amplitudes: we note however 
that our factorisation formula is an all-order result, and therefore will yield more
general results, once the appropriate ingredients are computed at the relevant
orders.

From the point of view of phenomenology, the most interesting applications of our 
results will concern the production of Higgs bosons in the gluon fusion channel, 
possibly in association with electroweak gauge bosons. We will however show explicitly 
that the formalism can be used also with (anti)quark initial states, and compare our 
results with existing calculations. In the gluon fusion channel, we will begin by 
showing how known properties of the single Higgs boson cross section emerge 
as a special case of our result. We next move on to multiple Higgs boson production, 
which has been the focus of much recent research. Beyond the leading-order 
result~\cite{Dawson:1998py}, we note that analytic expressions for the cross section 
are known only in the large top mass limit for Higgs pair production~\cite{deFlorian:2013jea,
deFlorian:2016uhr} and for triple Higgs boson production~\cite{deFlorian:2016sit}. 
In the case of Higgs pair production, leading order results with full top mass 
dependence were obtained in refs.~\cite{Glover:1987nx,Plehn:1996wb}, and 
numerical results have recently been presented at NLO~\cite{Borowka:2016ypz} 
(see also~\cite{Frederix:2014hta}); leading power threshold corrections have been 
considered in Ref.~\cite{Shao:2013bz}, and corrections to the large top mass 
approximation in Ref.~\cite{Grigo:2013rya}. In the case of triple Higgs boson 
production, numerical results with full top mass dependence at leading order 
and for real radiation at NLO were obtained in Ref.~\cite{Plehn:2005nk} and 
in Ref.~\cite{Maltoni:2014eza}, respectively. Associated production of electroweak 
bosons and Higgs bosons in the gluon fusion channel, discussed in 
Ref.~\cite{Hespel:2015zea}, also falls within the scope of our method, 
although we will not discuss it in detail here.

In this paper, we go beyond previous analytic results for Higgs boson pair 
production, by providing NLO corrections, up to NLP in the threshold variable, 
with full top mass dependence. As a further illustration and check, we demonstrate
consistency with known results for triple Higgs boson production in the large 
top mass limit~\cite{deFlorian:2016sit}, diphoton production~\cite{Gastmans:1990xh}, 
and the production of $W^+W^-$ pairs~\cite{Frixione:1993yp}. These results serve 
as an illustration of the method: we postpone a detailed phenomenological analysis, 
as well as applications to triple Higgs production with full top mass dependencce,
and to associated production of Higgs bosons with Z bosons, to future work. 
We note in passing that the present results also provide a strong consistency
check on all future NLO analytic computations of loop-induced processes of 
colour-singlet particles, which of course must agree with the simple factorised
expressions we derive at NLP.

The structure of our paper is as follows. In \secn{sec:review}, we briefly review 
the next-to-soft factorisation formula of Refs.~\cite{Bonocore:2015esa,
Bonocore:2016awd}, before describing how it can be extended for use in 
gluon-induced processes. In \secn{sec:result}, we derive an explicit expression
for the NLO cross-section of a colour-singlet final state in gluon fusion, valid up 
to NLP level. A similar result for the quark channel is derived in \secn{sec:quarks}. 
In \secn{sec:single}, we show how known results in single Higgs boson production 
are reproduced, before examining multiple Higgs boson production in \secn{sec:double}. 
Vector boson pair production is considered in \secn{sec:diphoton}. Finally, we discuss 
our results and future prospects in \secn{sec:conclude}. 


\section{NLP amplitude factorisation}
\label{sec:review}

In this section, we briefly review the results of Refs.~\cite{Bonocore:2015esa,
Bonocore:2016awd}, which derive a factorisation formula for QCD radiation up 
to NLP level, and we provide a generalisation of these results to the case of
external incoming gluons. Consider an amplitude with two incoming partons 
of momenta $p_1$ and $p_2$, and any number $N$ of final state colour singlet
particles, with momenta $p_i$, $3 \leq i \leq N + 2$. As described in detail in
Ref.~\cite{Bonocore:2016awd}, any such amplitude may be written in the
factorised form
\beq
  {\cal A}(\{ p_i \}) \, = \, \tilde{H} \left( \{p_i\}, n_1, n_2 \right) 
  \tilde{\cal S} \left( p_1, p_2 \right) \frac{\prod_{k=1}^2 J (p_k, n_k)}{\prod_{k=1}^2 
  \tilde{\cal J}(p_k, n_k)} \, .
\label{ampfac1}
\eeq
Here $\tilde{\cal S}(p_1, p_2)$ is a {\it next-to-soft function}, dressing the hard 
interaction with virtual exchanges of (next-to-)soft gluons, which couple to the 
external partons through the next-to-eikonal Feynman rules described in 
Refs.~\cite{Laenen:2008gt,Laenen:2010uz}. Associated with each hard parton 
is a {\it jet function} $J(p_i,n_i)$, which collects collinear singularities, and which 
depends on an auxiliary vector $n_i$. The {\it next-to-eikonal jet} $\, \tilde{\cal J}
(p_i,n_i)$ corrects for the double counting of contributions from gluons which 
are both (next-to-)soft and collinear, and finally the {\it hard function} $\tilde{H}
(\{p_i\}, n_1, n_2)$ is defined by matching to the amplitude on the left-hand side of 
\eq{ampfac1}, so that all dependence on the auxiliary vectors $\{n_i\}$ cancels out. 
If one ignores the presence of next-to-eikonal Feynman rules, \eq{ampfac1} 
reduces to the well-known soft-collinear factorisation formula, describing the 
dressing of a given hard interaction process with leading-power soft and collinear 
radiation (see, for example,~\cite{Dixon:2008gr}). The form of \eq{ampfac1},
however, is a crucial intermediate step in considering the emission of an 
additional gluon, of momentum $k^\mu$ and colour $a$. Up to NLP in this 
momentum, the resulting amplitude is given by~\cite{Bonocore:2016awd}
\beqa
  {\cal A}_\mu^{\, a} \left( \{p_i\}, k \right) & = & \sum_{l = 1}^2 \Bigg\{ 
  \Bigg[ \frac12 \, \frac{\tilde{\cal S}_\mu^{\, a} \left(p_1, p_2, k \right)}
  {\tilde{\cal S} (p_1, p_2)}
  + g_s \, {\bf T}_l^a \, G^\nu_{l, \mu} \, \frac{\partial}{\partial p_l^\nu} 
  + \frac{J_\mu^{\, a} \left(p_l, n_l, k \right)}{J(p_l, n_l)} 
  \label{NEfactor_nonabel} \\
  & & \qquad \quad - \, g_s \, {\bf T}_i^{\, a} \, G_{l, \mu}^\nu \,
  \frac{\partial}{\partial p_l^\nu} \log \left( \frac{J(p_l, n_l)}{\tilde{\cal J} (p_l, n_l)} 
  \right) \Bigg] {\cal A} \left( \{p_i\} \right) - {\cal A}^{\, a, \, \tilde{\cal J}_l}_\mu (\{p_i\},
  k) \Bigg\} \, , \nonumber 
\eeqa 
where $g_s$ is the QCD coupling\footnote{We absorb a factor $\mu^\e$, 
where $\mu$ is the dimensional regularisation scale, into the coupling for simplicity.}, 
${\bf T}_i^{\, a}$ a colour generator on line $i$ with adjoint index $a$, and we have 
introduced the tensor~\cite{Grammer:1973db} 
\beq 
  G^{\mu \nu}_l \, = \, \eta^{\mu \nu} - \frac{(2 p_l - k)^\nu}{2 p_l \cdot k - k^2} \, k^\mu \, .
\label{Gdef}
\eeq 
In addition to the functions already appearing in \eq{ampfac1}, \eq{NEfactor_nonabel} 
contains two more universal functions. First, the {\it radiative next-to-soft function}
$\tilde{\cal S}_\mu^{\, a}$ is a matrix element of next-to-eikonal Wilson lines directed 
along the directions of the incoming partons, like the virtual next-to-soft function 
$\tilde{\cal S}$, but with a single gluon present in the final state. Furthermore, 
\eq{NEfactor_nonabel} includes a {\it radiative jet function} $J_\mu^{\, a}$ collecting 
all contributions associated with the emission of a gluon from the $i^{\rm th}$ parton,
and enhanced by virtual collinear poles. This function was first introduced in the
context of abelian gauge theory in Ref.~\cite{DelDuca:1990gz}, and its definition was
recently generalised to non-abelian theories in Ref.~\cite{Bonocore:2016awd}. 
The radiative functions can be defined in terms of operator matrix elements, but
for our present NLO analysis, where radiative functions enter only at tree level,
a diagrammatic definition is sufficient.

The final term on the right-hand side of \eq{NEfactor_nonabel} is a subtraction 
term that removes any double counting of contributions occuring in both the 
radiative next-to-soft emission function, and in the radiative jet emission functions: 
it can be obtained simply by taking the next-to-soft limit of the radiative jet 
function. As was done in Ref.~\cite{Bonocore:2016awd}, \eq{NEfactor_nonabel} 
can be considerably simplified by using renormalisation group arguments and 
computing the right-hand side in the bare theory and with light-like reference 
vectors $n_i^2 = 0$ for the jets. With these choices, one can use the bare 
quantities
\beq 
  \tilde{\cal S} \left( p_1, p_2 \right) \, = \, J (p_i, n_i) \, = \, \tilde{\cal J} 
  \left(p_i, n_i \right) \, = \, 1 \, , \qquad \quad n_i^2 \, = \, 0 \, ,
\label{J0def}
\eeq
and the amplitude can be written as
\beqa
  {\cal A}_\mu^{\, a} \left( \{p_i\}, k \right) & = &
  \sum_{l = 1}^2 \Bigg\{ 
  \Bigg[ \frac12 \, \tilde{\cal S}_\mu^{\, a} \left( \{p_i\}, k \right)
  + g_s \, {\bf T}_l^{\, a} \, G^\nu_{l, \mu} \, \frac{\partial}{\partial p_l^\nu} 
  +J_\mu^{\, a} \left(p_l, n_l, k \right) \Bigg] {\cal A} \left(\{p_i\} \right) \notag \\
  & & \qquad - \, {\cal A}^{\, a, \, \tilde{\cal J}_l}_\mu \left( \{p_i\}, k \right) 
  \Bigg\} \, .
  \label{NEfactor2} 
\eeqa If we now focus on the NLO contributions to the cross section,
there is a further significant simplification: indeed, the
leading-order term in the next-to-soft emission function, $\tilde{\cal
  S}_{\mu, \, a}^{(1)}$ consists of single gluon emissions from the
hard incoming partons, and these contributions are completely
cancelled~\cite{Bonocore:2016awd} by the leading-order subtraction
term ${\cal A}^{(1), \tilde{\cal J}_l}_{\mu, \, a}$, leaving \beq
{\cal A}^{(1)}_{\mu, a} \left( \{p_i\}, k \right) \, = \, \sum_{l =
  1}^2 \Bigg[ g_s \, {\bf T}_{l, \, a} \, G^\nu_{l, \mu} \,
  \frac{\partial}{\partial p_l^\nu} + J^{(1)}_{\mu, \, a} \left( p_l,
  n_l, k \right) \Bigg] {\cal A}^{(0)} \left(\{p_i\}\right) \, ,
\label{NEfactor3} 
\eeq
which expresses the complete one-gluon radiative amplitude at NLO and 
NLP in terms of the Born amplitude. Note that in \eq{NEfactor3} the non-radiative 
amplitude and jet emission functions are understood as carrying implicit spin 
indices, depending on the identity of the particle species in each jet. 

\begin{figure}
\begin{center}
  \scalebox{0.6}{\includegraphics{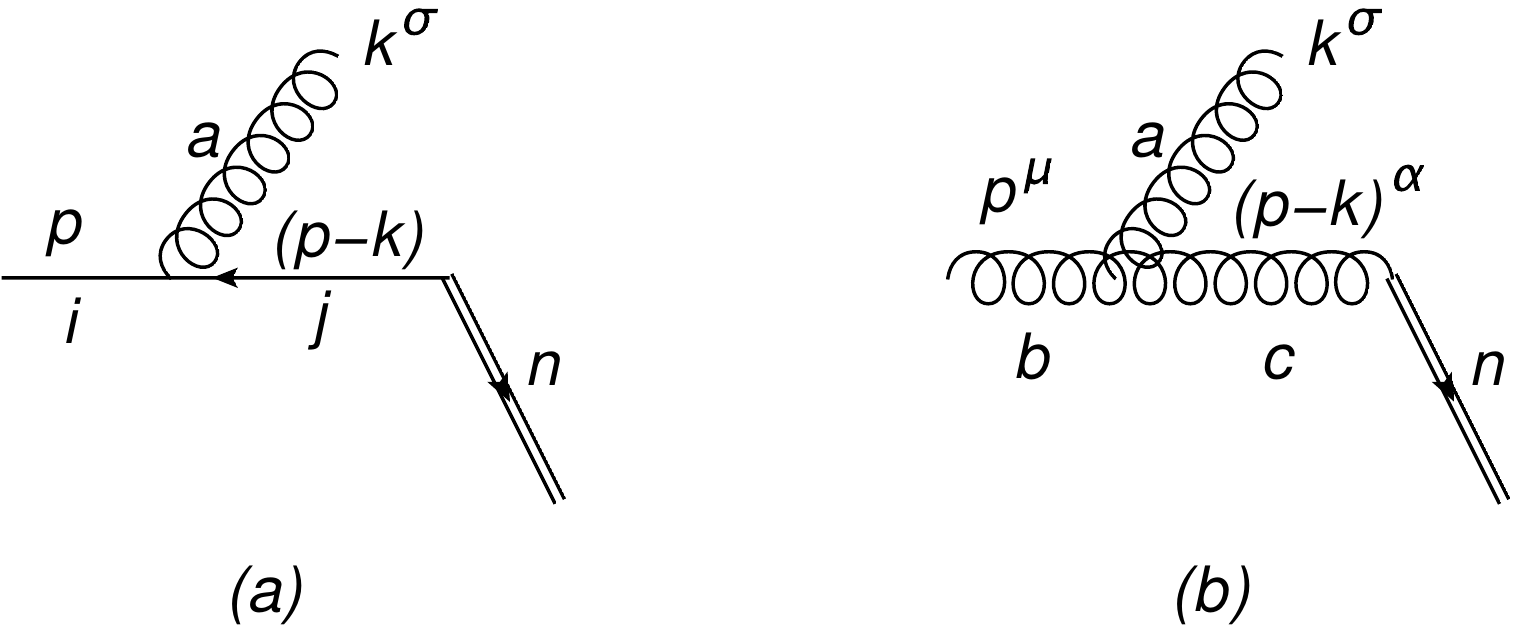}}
  \caption{Tree-level contribution to the radiative jet function for (a) an
  external quark; (b) an external gluon.}
\label{fig:Jmug}
\end{center}
\end{figure}

The quark radiative jet function at leading order is simply given by the emission of
a single gluon from the incoming (anti)quark~\cite{DelDuca:1990gz,Bonocore:2016awd}, 
as shown in Fig.~\ref{fig:Jmug}(a). Evaluating the diagram gives\footnote{Note that, 
by definition, the radiative jet function does not include the spinor wave function for 
the external quark line.}
\beq
  J_\mu^{\, a} (p, n, k) \, = \, g_s {\bf T}^{\, a} \left[ \frac{(2 p - k)_\mu}{2 p \cdot k}
  + \frac{{\rm i} k^\beta}{p \cdot k} \, S_{\beta \mu} \right] \, , \qquad
  S_{\beta \mu} \, = \, \frac{\rm i}{4} \left[ \gamma_\beta, \gamma_\mu \right] \, ,
\label{Jmuq} 
\eeq
where we have decomposed the result into spin-dependent and spin-independent
parts, introducing the generator $S_{\beta \mu}$ of Lorentz transformations on 
spinors. Note that at leading order the quark jet function is independent of the
auxiliary vector $n^\mu$, consistently with \eq{NEfactor3}, which represents a 
physical amplitude and cannot depend on $n$. For the gluon radiative jet function,
at leading order, we can simply use  a diagrammatic definition, analogous to the 
radiative quark jet, and shown in Fig.~\ref{fig:Jmug}(b).
Restoring explicit spin indices for the external gluon, we can write the result of this 
diagram as
\beq
  J_{\mu, \, \rho \sigma}^{\, a} (p, n, k) \, = \, g_s {\bf T}^{\, a} 
  \left[ \frac{(2 p - k)_\mu}{2 p \cdot k} \, \eta_{\rho \sigma} - 
  \frac{ {\rm i} k^\beta}{p \cdot k} \, M_{\beta \mu, \, \rho \sigma} \right] \, ,
\label{Jsigma2}
\eeq
where we have introduced the generator of Lorentz transformation
acting on vector fields,
\beq
  M_{\beta \mu, \, \rho \sigma} \, = \,  {\rm i} \Big( \eta_{\beta \rho} \eta_{\mu \sigma}
  - \eta_{\beta \sigma} \eta_{\mu \rho} \Big) \, .
\label{Mdef}
\eeq
Once again, we have decomposed the kinematic part into its spin-dependent 
and spin-independent parts (see for example~\cite{White:2014qia}). The colour 
operator for the gluon case can be explicitly interpreted as
\beq
  \left[ {\bf T}_{\, a} \right]_{bc}\, = \, {\rm i} f_{abc} \, .
\label{Tadef}
\eeq
Turning now to the derivative contribution to \eq{NEfactor3}, one may note that the
action of the projector $G^{\mu \nu}_l$ defined in \eq{Gdef}, up to NLP order, can 
be recast in terms of the orbital angular momentum of parton $l$. Indeed, to this 
accuracy
\beq
  G^\nu_{l, \mu} \frac{\partial}{\partial p_l^\nu} \, = \, \frac{k^\nu}{p_l \cdot k}
  \left[p_{l, \nu} \, \frac{\partial}{\partial p^\mu_l} - p_{l, \mu}\frac{\partial}{\partial 
  p_l^\nu} \right] \, = \, - \frac{{\rm i} k^\nu \, {\rm L}^{(l)}_{\nu \mu}}{p_i \cdot k} \, ,
\label{Ldef}
\eeq
where ${\rm L}^{(l)}_{\nu \mu}$ is the orbital angular momentum operator associated 
with the $l^{\rm th}$ parton. Using Eqs.~(\ref{Jsigma2}, \ref{Mdef}), we can now rewrite 
\eq{NEfactor3} in a unified notation for quarks and gluons, as
\beqa
  {\cal A}^{(1)}_{\mu, a} \left(\{p_i \}, k \right) & = &  
  \sum_{l = 1}^2  \, g_s \, {\bf T}_{l, a} \Bigg[
  \frac{(2 p_l - k)_\mu}{2 p_l \cdot k} - \frac{{\rm i} k^\nu}{p_l \cdot k}
  \left({\rm L}^{(l)}_{\nu \mu} + {\rm \Sigma}^{(l)}_{\nu \mu}\right) \Bigg]{\cal A}^{(0)} 
  \left(\{p_i\} \right) \,  \nonumber \\
  & = & \sum_{l = 1}^2 \, g_s \, {\bf T}_{l, a}
  \Bigg[ \frac{p_{l, \mu}}{p_l \cdot k} - \frac{{\rm i} k^\nu \, {\rm J}^{(l)}_{\nu \mu}}
  {p_l \cdot k} \Bigg] {\cal A}^{(0)} \left(\{p_i\} \right) \, ,
\label{NEfactor4} 
\eeqa
where in the first line ${\rm \Sigma}^{(l)}_{\nu \mu}$ is the spin angular momentum 
operator for parton $l$, in the relevant representation of the Lorentz group, acting 
as $- S^{(l)}_{\nu \mu}$ for spin one half, and as $M^{(l)}_{\nu \mu}$ for spin one, 
while ${\rm J}^{(l)}_{\nu \mu}$ is the total angular momentum operator. Furthermore,
in the second line, we have omitted the term proportional to $k_\mu$, which gives
a vanishing contribution when contracted with a physical polarisation vector for
the emitted gluon.

Equation~(\ref{NEfactor4}) is recognisable as the recently derived {\it next-to-soft
theorem}~\cite{Casali:2014xpa}, which mirrors a similar result derived in 
gravity~\cite{White:2011yy,Cachazo:2014fwa}. As noted, this formula encompasses 
both the quark and gluon cases, provided the spin operator is interpreted appropriately, 
validating our diagrammatic definition for the leading order gluon radiative jet function. 
For the NLO analysis performed in this paper, we could in fact have simply adopted 
\eq{NEfactor4} as the starting point for our following analysis; note, however, that
\eq{NEfactor_nonabel} and \eq{NEfactor2} are much more general results, applicable 
in principle to any order in perturbation theory.


\section{Colour-singlet particle production in the gluon channel}
\label{sec:result}

\begin{figure}
\begin{center}
  \scalebox{0.6}{\includegraphics{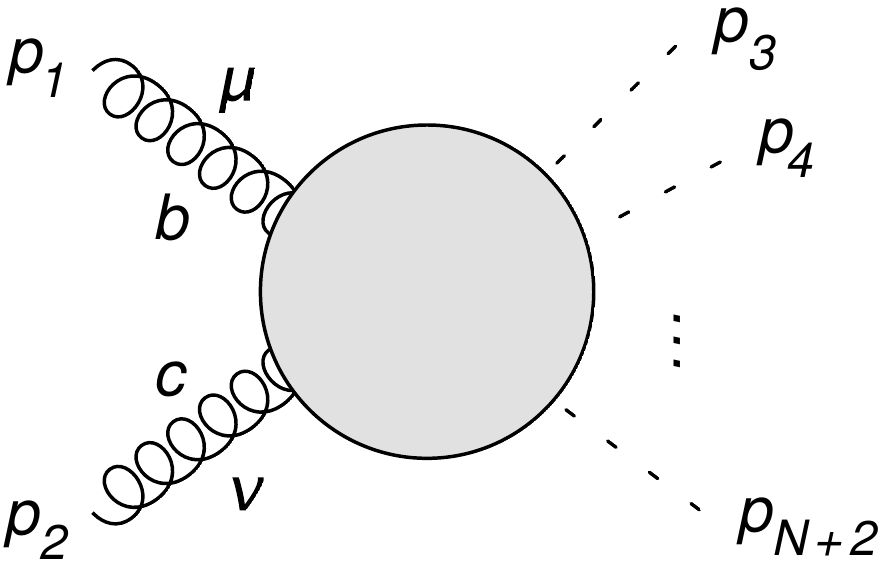}}
  \caption{The amplitude for the production of $N$ colour-singlet particles from a pair 
  of gluons, without final state QCD radiation.}
\label{fig:LOamp}
\end{center}
\end{figure}

In this section, we apply the result of \eq{NEfactor4} to obtain a general expression 
for the NLO cross-section for the production of $N$ colour-singlet particles near threshold. 
We begin by considering the gluon-induced process shown in Fig.~\ref{fig:LOamp}, while 
we will turn to the quark-induced process in \secn{sec:quarks}. At Born level, the momenta
introduced in Fig.~\ref{fig:LOamp} satisfy the leading-order momentum conservation condition
\beq
  \sum_{i = 1}^2 p_i^\mu \, = \, \sum_{i = 3}^{N + 2} p_i^\mu \, \equiv \, P^\mu \, ,
\label{momconLO}
\eeq
with the Born-level centre-of-mass energy squared given by $s \, = \, P^2$. Beyond 
Born level, we may define the dimensionless variable
\beq
  z \, = \, \frac{P^2}{s} \, ,
\label{zdef}
\eeq
which represents the fraction of the partonic centre-of-mass energy carried into the final 
state by all colour singlet particles. At leading order obviously $z = 1$; beyond leading 
order, additional real radiation may be emitted, in which case $0 \leq z \leq 1$, and $\xi 
\equiv 1 - z$ is a dimensionless threshold variable of the kind introduced in \eq{thresholddef}.
In particular, at NLO only a single gluon can be emitted, and all contributions up to NLP 
in the emitted momentum $k$ are captured by \eq{NEfactor4}. We can then use this to 
obtain a cross-section formula that is correct up to the first sub-leading order in $\xi$. 
To this end, it is useful to write the complete radiative amplitude (before contraction with 
external gluon polarisation vectors) as 
\beq
  {\cal A}^{\sigma, \, \mu \nu}_{\rm \, NLP} \, = \, {\cal A}_{\rm \, scal.}^{\sigma, \, \mu \nu}
  \, + \, {\cal A}_{\rm \, spin}^{\sigma, \, \mu \nu}
  \, + \, {\cal A}_{\rm \, orb.}^{\sigma, \, \mu \nu} \, ,
\label{Adecomp}
\eeq
where $\sigma$ is the Lorentz index of the emitted gluon, while $\mu$ and $\nu$ are the 
Lorentz indices of the incoming gluons, and, for simplicity, we have suppressed momentum 
dependence, colour indices and the superscript denoting the perturbative order. The 
three terms on the right-hand side correspond to the scalar, spin-dependent and orbital 
angular momentum terms in \eq{NEfactor4} respectively. The colour indices of the incoming 
gluons are displayed in Fig.~\ref{fig:LOamp}, and we note that by colour conservation the 
leading order amplitude must be proportional to $\delta^{bc}$. Following Eqs.~(\ref{Tadef}, 
\ref{NEfactor4}), one may write the scalar-like contribution to the amplitude as
\beq
  {\cal A}_{\rm \, scal.}^{\sigma, \, \mu \nu} \, = \, {\rm i} g_s f_{abc}
  \left[ \frac{ \left(2 p_1 - k \right)^\sigma}{2 p_1 \cdot k} - 
  \frac{\left(2 p_2 - k \right)^\sigma}{2 p_2 \cdot k} \right] {\cal A}^{\, \mu \nu} \, 
\label{AscaldiHiggs}
\eeq
where, as above, we omitted the superscript denoting the perturbative order 
for the Born amplitude ${\cal A}^{\, \mu \nu}$. Using Eqs.~(\ref{Mdef}, 
\ref{NEfactor4}), the spin-dependent contribution to the amplitude is given by
\beq
  {\cal A}_{\rm \, spin}^{\sigma, \, \mu \nu} \, = \, {\rm i} g_s f_{abc} \left[
  \frac{{\cal A}^{\, \alpha \nu}}{p_1 \cdot k} \left(k^\mu \delta_{\, \alpha}^\sigma
  - k_\alpha \eta^{\mu \sigma} \right) - \frac{{\cal A}^{\, \mu \alpha}}{p_2 \cdot k}
  \left(k^\nu \delta^\alpha_{\, \sigma} - k_\alpha \eta^{\nu \sigma} \right) \right] \, .
\label{AspindiHiggs}
\eeq
Finally, the orbital angular momentum contribution is 
\beq
  {\cal A}^{\sigma, \, \mu\nu}_{\rm \, orb.} \, = \, {\rm i} g_s f_{abc}
  \left[ G_1^{\alpha \sigma} \, \frac{\partial {\cal A}^{\, \mu \nu}}{\partial
  p_1^\alpha} \, - \, G_2^{\alpha \sigma} \, \frac{\partial {\cal A}^{\, \mu
  \nu}}{\partial p_2^\alpha} \right] \, .
\label{AderivdiHiggs}
\eeq
After including polarisation vectors for the two incoming gluons, the squared matrix 
element, accurate to NLP level and summed over polarisations and colours, is
\beqa
  \left| {\cal A}_{\rm \, NLP} \right|^2 & = & \sum_{\rm colours} \left( 
  {\cal A}^{\sigma_1, \, \mu_1 \nu_1}_{\rm \, scal.} + 
  {\cal A}^{\sigma_1, \, \mu_1 \nu_1}_{\rm \, spin} + 
  {\cal A}^{\sigma_1, \, \mu_1 \nu_1}_{\rm \, orb.} \right)^*
  {\cal P}_{\mu_1 \mu_2} (p_1, l_1) {\cal P}_{\nu_1 \nu_2} (p_2, l_2)
  {\cal P}_{\sigma_1 \sigma_2} (k, l_3)
  \nonumber \\
  & & \hspace{11mm} \times \, 
  \left({\cal A}^{\sigma_2, \, \mu_2 \nu_2}_{\rm \, scal.} + 
  {\cal A}^{\sigma_2, \, \mu_2 \nu_2}_{\rm \, spin} + 
  {\cal A}^{\sigma_2, \, \mu_2 \nu_2}_{\rm \, orb.} \right) \, ,
\label{A2NLPeff}
\eeqa
where we defined the polarisation sum
\beq
  {\cal P}_{\alpha \beta} (p,l) \, \equiv \, \sum_\lambda \e_\alpha^{\, (\lambda)} (p) \,
  \e^{\, (\lambda) *} _\beta(p) \, = \, - \eta_{\alpha \beta} + \frac{p_\alpha l_\beta +
  p_\beta l_\alpha}{p \cdot l} \, ,
\label{polsum}
\eeq
with $l$ is an arbitrary light-like reference vector used to define physical polarisation 
states, whose dependence must cancel in the final result. Alternatively, one could sum
over all polarisations, using ${\cal P}_{\alpha \beta} = - \eta_{\alpha \beta}$, and correct
for this by including external ghost contributions. Following this second approach, it
is fairly easy to conclude that ghost contributions vanish at NLP: indeed, final 
state ghost emission is suppressed by a power of the energy at amplitude level,
and thus contributes at NNLP at cross section level; furthermore, diagrams with a
ghost-antighost pair in the initial state do not couple directly to fermions or to the 
Higgs boson, and are strongly suppressed. These expectations are borne out by
a direct calculation, showing that all terms proportional to the vector $l^\mu$ in 
\eq{A2NLPeff} are beyond the required accuracy. We conclude that we can perform
a sum over all polarisation, so that \eq{A2NLPeff} simplifies to
\beq
  \left| {\cal A}_{\rm \, NLP} \right|^2 \, = \, \sum_{\rm colours}
  \left\{ \left| {\cal A}_{\rm \, scal.}^{\sigma, \, \mu \nu} \right|^2
  + 2 {\rm Re} \left[ \left( {\cal A}^{\sigma, \, \mu \nu}_{\rm \, spin}
  + {\cal A}^{\sigma, \, \mu \nu}_{\rm \, orb.} \right)^*
  {\cal A}_{\rm \, scal. \, \sigma, \, \mu \nu} \right] \right\} \, ,
\label{A2NLPeff2}
\eeq
where in the second term we need to keep only those terms which are leading power in 
the scalar part of the amplitude. It is straightforward to show that the first term on the 
right-hand side yields
\beq
  \sum_{\rm colours} \left|{\cal A}_{\rm \, scal.}^{\sigma, \, \mu \nu} \right|^2
  \, = \, 2 g_s^2 N_c \left( N_c^2 - 1 \right) \, \frac{p_1 \cdot p_2}{p_1 \cdot k \, p_2 \cdot k} \,
  |{\cal A}_{\, \mu \nu}|^2 \, ,
\label{scalsq}
\eeq
so that only the leading power term survives. For the gluon-initiated process we are 
considering in this section, the spin term in \eq{A2NLPeff2} can be shown to vanish upon 
summing over all polarisations. For example, the spin contribution from the first leg is
\beqa
  \sum_{\rm colours} 2 \, {\rm Re} \left[ {\cal A}_{\rm \, spin}^{\sigma, \, \mu \nu}
  {\cal A}^*_{\rm \, scal. \, \sigma, \, \mu \nu} \right]
  & = & - \, \frac{2 g_s^2 C_A \left( N_c^2 - 1 \right)}{p_1\cdot k} 
  \left(\frac{p_{1\sigma}}{p_1 \cdot k} - \frac{p_{2 \sigma}}{p_2 \cdot k} \right) \nonumber \\
  & & \, \times \, \Big( k^\mu \eta^{\sigma \alpha} - k^\alpha \eta^{\mu \sigma} \Big) \, 
  {\rm Re} \left[ {\cal A}_\alpha^{\, \nu}{\cal A}^*_{\, \mu \nu} \right] \, .
\label{spinsq}
\eeqa
The prefactor in the second line is antisymmetric under the interchange of $\alpha$ 
and $\mu$, and thus vanishes when contracted with the squared Born amplitude, 
which is symmetric; the same argument applies to the second incoming gluon. Note
that the argument applies also when the Born amplitude is loop induced, and thus
may acquire an imaginary part (as is the case here). The orbital angular momentum 
contributions give
\beqa
  \sum_{\rm colours} 2 \, {\rm Re} \left[{\cal A}_{\rm \, orb.}^{\sigma, \, \mu \nu}
  {\cal A}_{\rm \, scal. \, \sigma, \, \mu \nu} \right] & = & 
  - \, 2 g_s^2 N_c \left( N_c^2 - 1 \right) {\cal A}_{\, \mu\nu}
  \left[G_1^{\alpha \sigma} \, \frac{\partial {\cal A}^{\, \mu \nu}}{\partial p_1^\alpha}
  - G_2^{\alpha\sigma} \, \frac{\partial{\cal A}^{\, \mu \nu}}{\partial p_2^\alpha}
  \right] \nonumber \\
  & & \, \times \, \left(\frac{p_{1, \sigma}}{p_1 \cdot k} - \frac{p_{2, \sigma}}{p_2\cdot k}
  \right) \label{derivsq} \\[2mm]
  & = & \frac{2 g_s^2 N_c \left( N_c^2 - 1 \right) p_1 \cdot p_2}{p_1 \cdot k \, p_2 \cdot k} \,
  \Bigg[ \delta p_1^\alpha \, \frac{\partial}{\partial p_1^\alpha}
  + \delta p_2^\alpha \, \frac{\partial}{\partial p_2^\alpha} \Bigg]
  \left| {\cal A}_{\mu \nu} \right|^2 \, , \nonumber
\eeqa
where we defined
\beq
  \delta p_1^\alpha \, = \, - \frac12 \left( \frac{p_2 \cdot k}{p_1 \cdot p_2} \, p_1^\alpha
  - \frac{p_1 \cdot k}{p_1 \cdot p_2} \, p_2^\alpha + k^\alpha \right) , \, \, \, 
  \delta p_2^\alpha \, = \, - \frac12 \left( \frac{p_1 \cdot k}{p_1 \cdot p_2} \, p_2^\alpha
  - \frac{p_2 \cdot k}{p_1 \cdot p_2} \, p_1^\alpha + k^\alpha \right) .
\label{deltapdef}
\eeq
Note that these shifts are proportional to the soft momentum $k$ and transverse to their 
respective momenta, $p_i \cdot \delta p_i = 0$. This second property follows from the 
fact that the $i^{\rm th}$ momentum shift is derived from the orbital angular momentum
 operator of \eq{Ldef}, which generates an infinitesimal Lorentz transformation transverse
 to the momentum $p_i$. Combining \eq{derivsq} with \eq{scalsq}, we can write
\beq
  \left| {\cal A}_{\rm NLP} \right|^2 \, = \, \frac{2 g_s^2 N_c \left(N_c^2 - 1 \right) 
  p_1\cdot p_2} {p_1 \cdot k \, p_2 \cdot k} \left| {\cal A}_{\, \mu \nu}
  \left(p_1 + \delta p_1, p_2 + \delta p_2 \right) \right|^2 \, .
\label{ANLPshift}
\eeq
\eq{ANLPshift} is a focal point of this paper: it shows that all NLP contributions 
to the NLO squared matrix element for the production of an arbitrary colour-singlet
final state can be absorbed into a shift in the kinematics of the Born contribution. 
Corrections to this shifting procedure involve terms at least quadratic in $\delta p_i$, 
and thus beyond the NLP approximation. In \secn{sec:quarks}, we will show that 
the same property is shared by quark-initiated processes. Note that \eq{ANLPshift} 
is fully differential in final state momenta, and can be applied to generate distributions 
valid at NLO and to NLP accuracy. On the other hand, using simple properties of 
phase space, one can also derive a similarly simple expression for the inclusive 
cross section. In order to do so, note that the effect of the required momentum 
shifts on the partonic centre-of-mass energy is given by a simple rescaling. Indeed,
\beq
  s \, \rightarrow \, \left( p_1 + p_2 + \delta p_1 + \delta p_2 \right)^2 \, = \, 
  s + 2 \left( \delta p_1 + \delta p_2 \right) \cdot (p_1 + p_2) \, .
\label{sshift}
\eeq
Substituting the definitions of \eq{deltapdef} in \eq{sshift}, and using \eq{zdef}, 
together with the NLO momentum conservation condition
\beq
  p_1 + p_2 \, = \, k + P \, , 
\label{NLOmomcon}
\eeq
it is easy to show that \eq{sshift} can be written simply as
\beq
  s \, \rightarrow \, z s \, . 
\label{sshift2}
\eeq
To construct the partonic cross-section, we must now introduce the appropriate 
factors to average over initial state colours and spins, integrate over the 
($N$ + 1)-body final state phase space, and include the flux factor. We find
\beq
  \hat{\sigma}_{\rm NLP}^{(gg)} \, = \, \frac{1}{(d - 2)^2 \left( N_c^2 - 1 \right)^2}
  \frac{1}{2 s} \int d \Phi_{N + 1} \left(P + k; p_3, \ldots, p_{N + 2}, k \right)
  \left| {\cal A}_{\rm NLP} \right|^2 \, ,
\label{sigmaNLPdef}
\eeq
where 
\beq
  d \Phi_{n} \left(Q; \{q_i\} \right) \, = \, (2 \pi)^d \, \delta^{(d)} 
  \Big(Q - \sum_{i = 1}^n q_i \Big) \prod_{i=1}^n \frac{d^{d - 1} 
  \vec{q}_i}{(2 \pi)^{d - 1} \, 2 E_i}
\label{PSdef}
\eeq
denotes the $n$-body Lorentz-invariant phase space for a process with total 
final state momentum $Q = \sum_i q_i$, and $q_i^\mu = (E_i, \vec{q}_i)$ in
a suitable frame. For the phase space, we may use the well-known result
\beq
  \int d \Phi_{N + 1} \left(P + k; p_3, \ldots p_{N + 1}, k \right) \, = \, 
  \frac{1}{2 \pi} \int d P^2 \, d \Phi_2 \left(P + k; P, k \right) \, 
  d \Phi_N \left(P; p_3, \ldots p_{N + 2} \right) \, ,
\label{phasespacefac}
\eeq
factorising the phase space of the $N$ colour-singlet particles from a two-body 
phase space involving the total momentum of the colourless system, and the 
additional gluon momentum $k$. The latter can be written more explicitly by 
parametrising
\beq
  p_1 \, = \, \frac{\sqrt{s}}{2} \left(1, 0, \ldots, 0, 1 \right) \, , \quad
  p_2 \, = \, \frac{\sqrt{s}}{2} \left(1, 0, \ldots, 0, - 1 \right) \, , \quad
  k \, = \, \frac{(1 - z) \sqrt{s}}{2} \left(1, 0, \ldots, \sin \chi, \cos \chi \right) \, .
\label{momparam}
\eeq
Introducing the variable
\beq
  y \, = \, \frac{1 + \cos \chi}{2} \, ,
\label{ydef}
\eeq
one then finds (see for example Ref.~\cite{Laenen:2010uz} for a recent 
derivation)
\beq
  d \Phi_2 \left(P + k; P, k \right) \, = \, \left( \frac{4 \pi}{s} \right)^\e
  \frac{1}{8 \pi \, \Gamma(1 - \e)} \, (1 - z)^{1 - 2 \e} \, 
  \Big[ y (1 - y) \Big]^{- \e} d y \, .
\label{dPS2}
\eeq
Using \eq{dPS2} together with Eqs.~(\ref{zdef}, \ref{ANLPshift}) in 
\eq{sigmaNLPdef}, one then finds
\beqa
  \frac{d \hat{\sigma}_{\rm NLP}^{(gg)}}{dz} & = &
  \frac{g_s^2 N_c}{8 (d - 2)^2 \pi^2 \left( N_c^2 - 1 \right) 
  \Gamma(1 - \e) \, s} \left( \frac{4 \pi \mu^2}{s} \right)^\e
  \int_0^1 dy \Big[ y (1 - y) \Big]^{- \e - 1}(1 - z)^{ - 1 - 2 \e} 
  \nonumber \\ & &
  \times \, \int d \Phi_N^{(z)} \, \Big| {\cal A}_{\mu\nu}
  (p_1 + \delta p_1, p_2 + \delta p_2) \Big|^2 \, ,
\label{dsigmadz}
\eeqa
where we reinstated the explicit dependence on the dimensional regularisation 
scale $\mu$, and we denoted by $d \Phi_N^{(z)}$ the phase space for $N$ 
(colour-singlet) particles with a partonic centre-of-mass energy shifted according 
to \eq{sshift2}. We may easily rewrite this result in terms of the leading-order cross 
section with shifted kinematics, which is given by
\beq
  \sigma^{(gg)}_{\rm Born} \left( z s \right) \, = \, \frac{1}{2 (d - 2)^2 
  \left( N_c^2 - 1 \right) \, z s} \, \int d \Phi_N^{(z)} \,
  \Big|{\cal A}_{\mu\nu} \left( p_1 + \delta p_1, p_2 + \delta p_2 \right) \Big|^2 \, .
\label{sigma0def}
\eeq
This leads us to our second central result: a simple factorised expression for the 
inclusive cross section, valid at NLO and NLP for the production of a generic
colour-singlet system, which can be written as
\beq
  \frac{d \hat{\sigma}_{\rm NLP}^{(gg)}}{dz} \, = \, C_A \, K_{\rm NLP} 
  \left( z, \e \right) \hat{\sigma}_{\rm Born}^{(gg)} \left(z s \right) \, ,
\label{simpfact}
\eeq
where the next-to-leading power $K$ factor is easily computed, with the result
\beqa
  K_{\rm NLP} \left( z, \e \right) & = & \frac{\alpha_s}{\pi}
  \left( \frac{4 \pi \mu^2}{s} \right)^\e
  z \, (1 - z)^{- 1 - 2 \e} \frac{\Gamma^2(- \e)}{\Gamma(- 2 \e)
  \Gamma(1 - \e)} \nonumber \\
  & = & \frac{\alpha_s}{\pi} \left( \frac{\overline{\mu}^2}
  {s} \right)^\e \left[ \frac{ 2 - 2 {\cal D}_0 (z) }{\e}
  + 4 {\cal D}_1(z) - 4 \log(1 - z) \right] \, .
\label{dsigmadz2}
\eeqa
In the second line of \eq{dsigmadz2} we expanded the result to NLP in 
$(1 - z)$ and to finite order in $\e$, and we introduced the $\overline{\rm MS}$ 
scale $\overline{\mu}^2 = \mu^2 \, {\rm e}^{\, \ln (4 \pi) - \gamma_E}$ and the 
plus distributions
\beq
  {\cal D}_i (z) \, \equiv \, \left( \frac{\log^i (1 - z)}{1 - z} \right)_+ \, .
\label{plusdef}
\eeq
Eqs.~(\ref{simpfact}, \ref{dsigmadz2}) show explicitly that the NLO K-factor 
for the production of $N$ colour-singlet particles in the gluon channel is 
simple and universal, up to next-to-leading power in the threshold variable. 
This is a powerful constraint, and we will discuss some specific examples 
in the following sections. First, however, we consider an analogous formula 
in the quark channel.


\section{Colour-singlet particle production in the quark channel}
\label{sec:quarks}

\begin{figure}
\begin{center}
  \scalebox{0.6}{\includegraphics{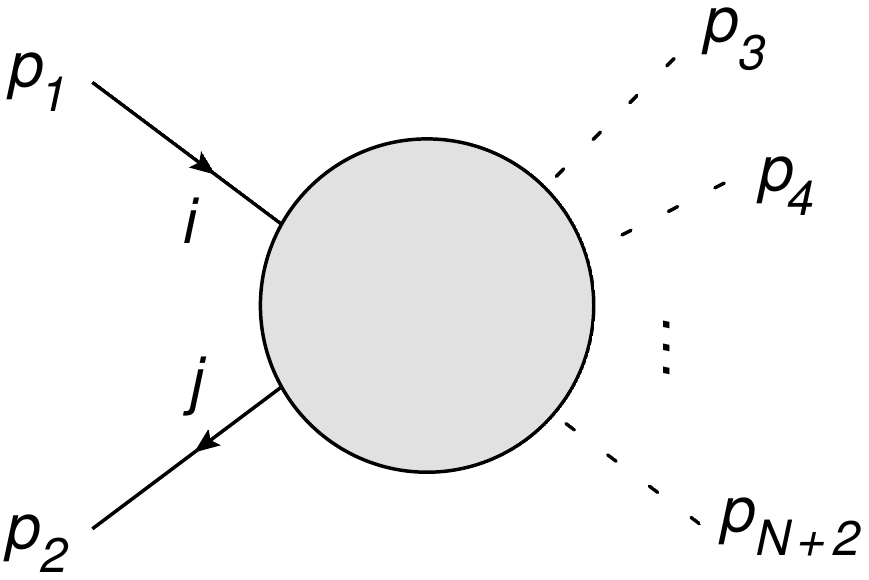}}
  \caption{The amplitude for the production of $N$ colour-singlet particles from a 
  quark-antiquark pair, without final state QCD radiation.}
\label{fig:LOamp2}
\end{center}
\end{figure}

In the previous section, we have derived an explicit universal K-factor for
multiple colour-singlet particle production in the gluon-gluon channel. In 
this section, we consider the cross section for quark-induced production 
of colour-singlet particles, and show that an identical result holds, up to
a trivial replacement of colour factors. The universality of the result is not
obvious from the outset, and it comes about through an interesting reshuffling
of the contributions of spin and angular momentum operators, as compared
to the gluon-induced process. We take the leading order process shown in 
Fig.~\ref{fig:LOamp2}, and consider the radiation of an additional gluon from 
the incoming quark and antiquark lines.
One may write the LO amplitude as
\beq
  A \left( \{ p_i \} \right) \, = \, \delta_{ij} \, \bar{v}(p_2) \, {\cal A} \left( \{ p_i \} \right) 
  u(p_1) \, ,
\label{A0quark}
\eeq
where $i$, $j$ are the colour indices of the incoming quark and antiquark, and the 
factor ${\cal A} (\{ p_i \})$, matrix-valued in spinor space, is the quantity entering 
\eq{NEfactor3}, namely the leading-order amplitude with external wave functions 
removed. Following the procedure adopted in the gluon case, we may decompose 
the NLO amplitude, before contraction with external spinors, according to
\beq
  {\cal A}^{\sigma}_{\rm \, NLP} \, = \, {\cal A}_{\rm \, scal.}^\sigma + 
  {\cal A}_{\rm \, spin}^\sigma + {\cal A}_{\rm \, orb.}^\sigma \, ,
\label{Adecomp2}
\eeq
where the three terms on the right-hand side denote the scalar-like, spin, and 
orbital angular momentum contributions, and we have suppressed spinor indices 
(as well as color labels) for brevity. For the scalar and orbital angular momentum 
contributions, which do not depend explicitly on the spin (apart from replacing 
the vector indices on the leading-order amplitude with spinor indices), the 
arguments of the previous section may be repeated, and one may 
write\footnote{Following the convention of \eq{A0quark}, we do not include 
the colour generator $t^a_{ji}$ in the fundamental representation in the definition
of the stripped amplitude ${\cal A}$.}
\beq
  {\cal A}^\sigma_{\rm \, scal.} + {\cal A}^\sigma_{\rm \, orb.} \, = \, 
  g_s \left( \frac{p_1^\sigma}{p_1 \cdot k} - \frac{p_2^\sigma}{p_2 \cdot k} \right)
  {\cal A} \left( p_1 + \delta p_1, p_2 + \delta p_2 \right) \, ,
\label{Ascalorb} 
\eeq
where the momentum shifts are defined in \eq{deltapdef}. Including the (anti)quark
wave functions and performing color and spin sums, we then find
\beqa
  && \hspace{-6mm}
  \left| A_{\rm \, NLP}^\sigma \right|^2_{\rm \, \, scal. \, + \, orb.} \, = \, 
  g_s^2 N_c C_F \, \frac{2 p_1 \cdot p_2}{p_1 \cdot k \, p_2 \cdot k} \,
  {\rm Tr}\left[\slsh{p_1}{\cal A} \left( p_1 + \delta p_1, p_2 + \delta p_2 \right)
  \slsh{p}_2 {\cal A}^\dagger \left( p_1 + \delta p_1, p_2 + \delta p_2 \right) 
  \right] \nonumber \\
  && \hspace{5mm}
  = \, g_s^2 N_c C_F \, \frac{s^2}{p_1 \cdot k \, p_2 \cdot k} \, 
  {\rm Tr} \left[ \slsh{n_1} {\cal A} \left( p_1 + \delta p_1, p_2 + \delta p_2 \right)
  \slsh{n}_2 {\cal A}^\dagger \left( p_1 + \delta p_1, p_2 + \delta p_2 \right) \right] \, ,
\label{Ascalorb2}
\eeqa
where in the second line we have introduced the dimensionless vectors
\beq
  n_i^\mu \, = \, \frac{p_i^\mu}{\sqrt{s}} \, \qquad \quad i = 1,2 \, .
\label{nidef}
\eeq
By comparing \eq{Ascalorb2} with its LO counterpart,
\beq
  \left| A(p_1, p_2) \right| \, = \, N_c \, s \, {\rm Tr} \left[
  \slsh{n}_1 {\cal A} (p_1, p_2) {\slsh n}_2 {\cal A}^{\dagger}
  (p_1, p_2) \right] \, ,
\label{A0sq}
\eeq
and using \eq{sshift2} we may promote the momentum shift in \eq{Ascalorb2} 
to apply to the entire squared amplitude. This leads to
\beq
  \left| A_{\rm \, NLP}^\sigma \right|^2_{\rm \, \, scal. \, + \, orb.} \, = \, \frac{g_s^2
  C_F}{z} \, \frac{s}{p_1 \cdot k \, p_2 \cdot k} \, 
  \left| A \left( p_1 + \delta p_1, p_2 + \delta p_2 \right) \right|^2 \, .
\label{ANLP2}
\eeq
Note the close resemblance of \eq{ANLP2} and \eq{ANLPshift}: they differ only
by the color factor and a rescaling by a factor of $z$. We must still, however,
add to \eq{ANLP2} the interference between the spin-dependent  part of the 
NLO amplitude, and the eikonal amplitude.  In the gluon case, this turned out 
to vanish in Feynman gauge, upon summing over all gluon polarisations, which 
was allowed at NLP accuracy. For an incoming fermion, we find that the spin 
contribution does not vanish, and indeed it precisely compensates for the
$z$ rescaling observed in \eq{ANLP2}, recovering the universality of the result.

\begin{figure}
\begin{center}
  \scalebox{0.6}{\includegraphics{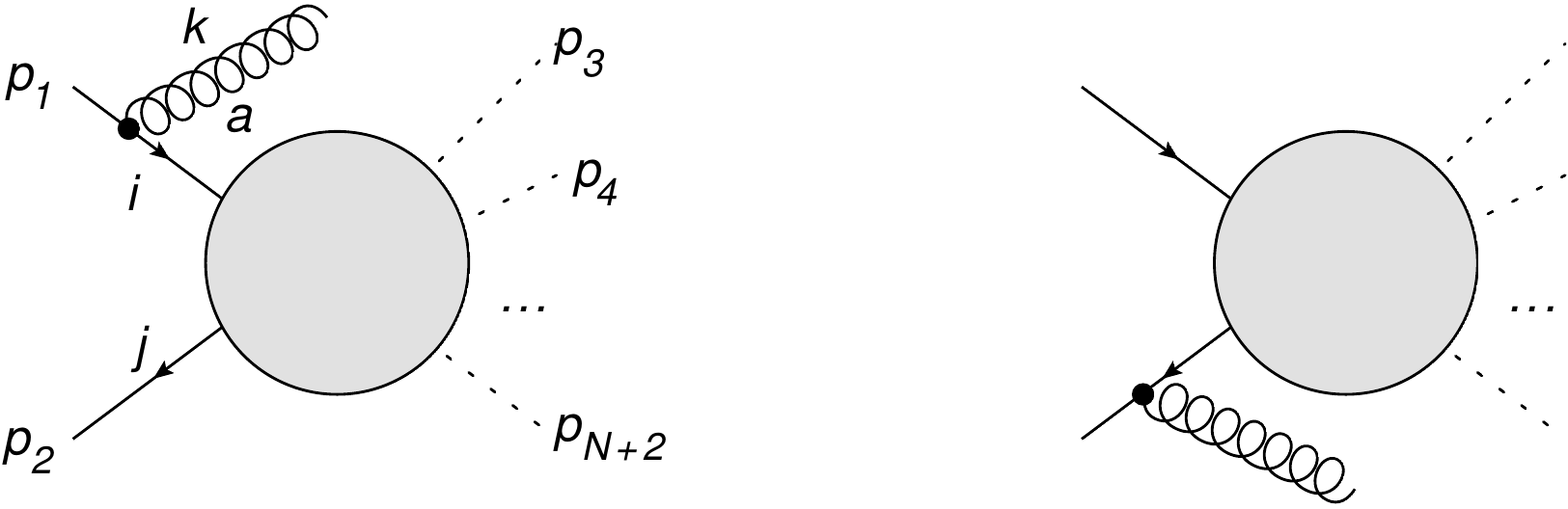}}
  \caption{Diagrams contributing to the spin-dependent part of the NLO
    cross-section, where $\bullet$ denotes the magnetic moment coupling of
    the gluon to the spin of the quark.}
\label{fig:NLOspindiags}
\end{center}
\end{figure}

The spin-dependent part of the NLO amplitude is given by the diagrams of 
Fig.~\ref{fig:NLOspindiags}, which evaluate to
\beqa
  \e_\sigma(k) \, \, t^a_{ji}\, \bar{v}(p_2) \, {\cal A}^\sigma_{\rm \, spin} \, u(p_1)
  & = & - \, {\rm i} \, g_s\, t^a_{ji}\,k_\beta \, \e_\sigma(k) \, \bar{v}(p_2)
  \left[  \frac{{\cal A} \, \Sigma^{\beta \sigma}}{p_1 \cdot k}
  + \frac{\Sigma^{\beta \sigma} {\cal A}}{p_2 \cdot k} \right] u(p_1)
  \nonumber \\ & = & 
  - \, g_s \, t^{a}_{ji} \, \e_\sigma(k) \, \bar{v}(p_2) \left[
  \frac{{\cal A} \, \slsh{k} \, \gamma^\sigma}{2 p_1 \cdot k} - 
  \frac{\gamma^\sigma \slsh{k} \, {\cal A}}{2 p_2 \cdot k}
  \right] u(p_1) \, .
\label{Aspinquark}
\eeqa
\hspace{-1.5mm} In the second line of \eq{Aspinquark}, we anticommuted Dirac 
matrices and used the physical polarisation condition $k^\sigma \e_\sigma(k) = 0$ 
to write the result in a form which will be more convenient in what follows. Up to 
NLP accuracy in the squared amplitude, we only need to consider the interference 
of \eq{Aspinquark} with the (leading power) scalar part of the NLO amplitude. 
Furthermore, since we are considering the emission of a single gluon, we can 
sum over all polarisations, rather than restricting to physical polarisations only.
The relevant contribution to the squared matrix element is then
\beq
  \sum_{\rm colours} 2\, {\rm Re} \left[ A^\dagger_{\rm \, scal.} \, 
  A_{\rm \, spin} \right]_{\rm \, NLP} \, = \, - \, \frac{g_s^2 N_c \, C_F}{p_1 
  \cdot k \, p_2 \cdot k} \, {\rm Tr} \Big[ \slsh{p}_2 \Big({\cal A} \slsh{k} \slsh{p}_2
  + \slsh{p}_1 \slsh{k} {\cal A} \Big) \slsh{p}_1 {\cal A}^{\dagger} \Big] \, .
\label{interference}
\eeq
To simplify this further, we may expand the emitted gluon momentum in
the Sudakov decomposition
\beq
  k^\sigma \, = \, \frac{p_2 \cdot k}{p_1 \cdot p_2} \, p_1^\sigma
  + \frac{p_1 \cdot k}{p_1 \cdot p_2} \, p_2^\sigma + k_T^\sigma \, , \qquad
  k_T \cdot p_1 \, = \, k_T \cdot p_2 \, = \, 0 \, .
\label{Sudakov}
\eeq
We then observe that, to linear order in $k^\mu$, the Dirac trace in \eq{Sudakov}
cannot depend on $k_T$. Indeed, one easily finds
\beq
  \sum_{\rm colours} 2\, {\rm Re} \left[ A^\dagger_{\rm \, scal.} \, 
  A_{\rm \, spin} \right]_{\rm \, NLP} \, = \, - \, g_s^2 N_c \, C_F \,
  \frac{2 p_1 \cdot p_2}{p_1 \cdot k \, p_2 \cdot k} \,
  \frac{k \cdot (p_1 + p_2)}{p_1 \cdot p_2} \, \left| {\cal A} (p_1,p_2) \right|^2 \, . 
\label{interference2}
\eeq
By comparing with the squared scalar part of the amplitude
\beq
  \sum_{\rm colours} A^\dagger_{\rm \, scal.} \, A_{\rm \, scal.}
  \, = \, g_s^2 N_c \, C_F \, \frac{2 p_1 \cdot p_2}{p_1 \cdot k \, p_2 \cdot k}
  \left| {\cal A} (p_1, p_2) \right|^2 \, ,
\label{scalsqquark}
\eeq
we see that the spin-dependent contribution to the squared amplitude can be 
obtained from the part which is leading power in the gluon momentum, simply
through rescaling by the factor
\beq
  - \frac{k \cdot (p_1 + p_2)}{p_1 \cdot p_2} \, = \, - \, (1 - z) \, ,
\label{NEfac}
\eeq
where we have used the momentum parametrisation of \eq{momparam}. 

Combining \eq{ANLP2} with \eq{scalsqquark}, we see that the rescaling factors
cancel at NLP in $(1 - z)$. Indeed one may write
\beq
  \left| A_{\rm \, NLP} \right|^2 \, = \, g_s^2 \, C_F \frac{s}{p_1 \cdot k \, 
  p_2\cdot k} \Bigg\{ \frac{ \left| A (p_1 + \delta p_1, p_2 + \delta p_2) \right|^2}{z} 
  - \left| A (p_1, p_2) \right|^2 (1 - z) \Bigg\} \, .
\label{ANLP3bis}
\eeq
Expanding now in powers of $(1 - z)$, one gets  to first order
\beqa
  \left| A_{\rm NLP} \right|^2 & = & g_s^2 \, C_F \frac{s}{p_1 \cdot k \, 
  p_2\cdot k} \Bigg\{ \left| A (p_1 + \delta p_1, p_2 + \delta p_2) \right|^2 \nonumber \\ 
  & & + \, \, \Big( \left| A (p_1 + \delta p_1, p_2 + \delta p_2) \right|^2 
  - \left| A (p_1, p_2) \right|^2 \Big) (1 - z) \Bigg\} \, ,
\label{ANLP3tris}
\eeqa
and one observes that the second line is effectively ${\cal O} (1 - z)^2$. We find then
\beq
  \left| A_{\rm \, NLP} \right|^2 \, = \, g_s^2 \, C_F \, \frac{s}{p_1 \cdot k \, 
  p_2\cdot k} \left| A (p_1 + \delta p_1, p_2 + \delta p_2) \right|^2 \, ,
\label{ANLP3}
\eeq
which is precisely analogous to \eq{ANLPshift}, except for the replacement of the
colour factor, which here is associated with the fundamental rather than adjoint 
representation of the gauge group. Once again, at NLP, \eq{ANLP3} can be used
in a fully differential implementation for the final state kinematics.

Having obtained \eq{ANLP3}, one may form the cross-section by integrating with 
the $(N + 1)$-body phase space, exactly as was done in the gluon case. One 
finds then 
\beq
  \frac{d \hat{\sigma}_{\rm \, NLP}^{(qq)}}{dz} \, = \, C_F \, K_{\rm NLP} 
  \left( z, \e \right) \hat{\sigma}_{\rm Born}^{(qq)} \left(z s \right) \, ,
\label{simpfact2}
\eeq
with {\it the same} factor $K_{\rm NLP} \left( z, \e \right)$, given in \eq{dsigmadz2}.

A first check on this result is that it reproduces the NLO K-factor for Drell-Yan 
production of a vector boson of invariant mass $Q^2$, where one has
\beq
  z \, = \, \frac{Q^2}{s} \, .
\label{zdefDY}
\eeq
In this case, as for any $2 \to 1$ process, the LO partonic cross section has support
only on the partonic threshold: for Drell-Yan production,
\beq
  \sigma_{\rm Born}^{(qq)} (s) \, \propto \, \delta (Q^2 - s) \, = \, \frac{1}{s} \,
  \delta \left(\frac{Q^2}{s} - 1 \right) \, ,
\label{DYLO}
\eeq
so that the LO cross section with shifted kinematics is 
\beq
  \sigma_{\rm Born}^{(qq)} (z s) \, \propto \, \delta \left( Q^2 - z s \right) \, = \, 
  \frac{1}{s} \, \delta \left( \frac{Q^2}{s} - z \right) \, . 
\label{DYLOshift}
\eeq
The delta-function imposes the correct definition of the threshold variable at 
NLO, while the rest of the cross section is unaffected by the shift in kinematics. 
To compare with standard results, we must note that the $\overline{\rm MS}$ 
scale $\overline{\mu}^2$ is usually set equal to the final state invariant mass
$Q^2$. To this end, one may write
\beq
  \left( \frac{\mu^2}{s} \right)^\e \, = \, \left( \frac{\mu^2}{Q^2} \right)^\e
  \left( \frac{Q^2}{s} \right)^\e \rightarrow z^\e \, ,
\label{scales}
\eeq
so that \eq{simpfact} becomes
\beqa
  \frac{d \hat{\sigma}_{\rm \, NLP}^{q q}}{d z} & = & \frac{\alpha_s}{\pi} \, C_F \,
  z^{\e} \left[ \frac{2 - 2 {\cal D}_0 (z)}{\e} + 4 {\cal D}_1 (z)  -
  4 \log(1 - z) \right] \sigma_{\rm Born}^{(qq)} (z s) \nonumber \\
  & = & \frac{\alpha_s}{4 \pi} \, C_F \, \left[ \frac{8 - 8 {\cal D}_0 (z)}{\e}
  + 16 {\cal D}_1 (z) - 16 \log(1 - z )+ 8 \right] \, \sigma_{\rm Born}^{(qq)} (z s)  \, ,
\label{Kfacquark2}
\eeqa
which precisely agrees with the well-known results quoted for example in
Refs.~\cite{Hamberg:1991np,Laenen:2010uz}.


\section{Single Higgs boson production via gluon fusion}
\label{sec:single}

Having presented our results for both quark- and gluon-induced colour-singlet 
particle production, we now examine a first significant application of the 
gluon result, \eq{dsigmadz2}: the single production of Higgs bosons in the 
gluon fusion channel. As is well known, this is is the principal production mode 
for Higgs bosons at the LHC, and higher-order QCD corrections have been 
studied in great detail and with great efforts in recent years. In the effective 
field theory with the top quark integrated out, they have been calculated up 
to N$^3$LO in perturbation theory~\cite{Dawson:1991zj,Anastasiou:2002yz,
Harlander:2002wh,Ravindran:2003um,Anastasiou:2013mca,Anastasiou:2015ema,
Anastasiou:2014lda,Anastasiou:2014vaa,Anastasiou:2013srw}. Top-mass 
effects are know exactly at NLO~\cite{Spira:1995rr}, and have been studied 
at NNLO as a power expansion in $m_h^2/m_t^2$~\cite{Harlander:2009mq,
Pak:2009dg}. Here we will see how the intricate top mass dependence at NLO
simplifies considerably in the threshold region, including NLP corrections.

\begin{figure}
\begin{center}
  \scalebox{0.7}{\includegraphics{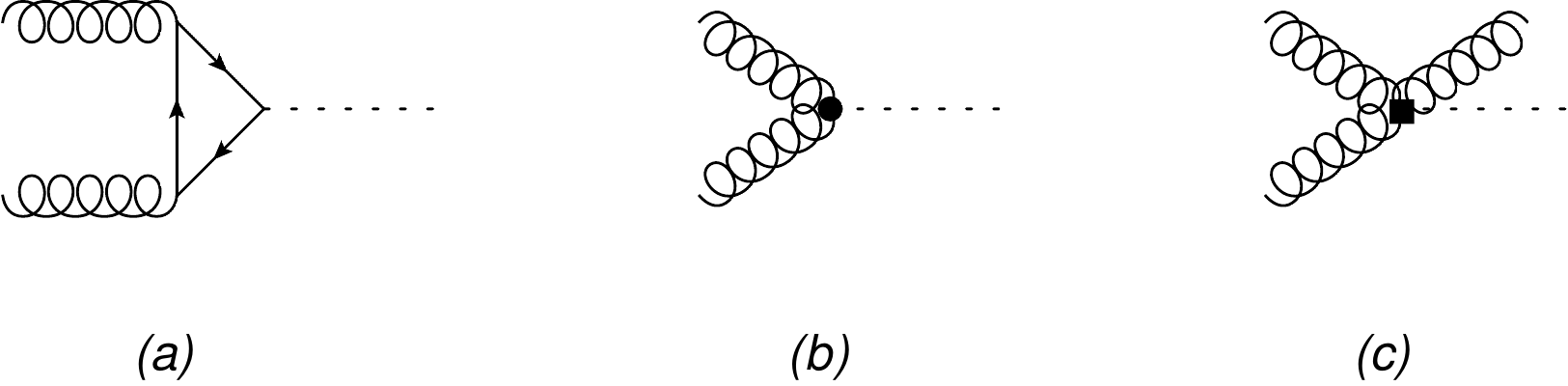}}
  \caption{(a) Leading order diagram for the production of a Higgs boson via 
    gluon fusion; (b) Contact interaction in the large top mass limit; (c) Contact 
    interaction for radiation of an extra gluon.}
\label{fig:singleHiggs}
\end{center}
\end{figure}

At leading order, the incoming gluons couple to the Higgs boson via a
top-quark loop, as shown in Fig.~\ref{fig:singleHiggs}(a).
The leading order cross-section for this process (see for example~\cite{Dawson:1991zj})
can be written as
\beq
  \sigma^{h}_{\rm \, Born} (s) \, = \, \frac{\alpha_s^2}{\pi} \, 
  \frac{m_h^2}{576 v^2} \, (1 + \e) \, F(\tau, \e) \, \delta ( s - m_h^2) \, ,
\label{sigma0Higgs}
\eeq
where $m_h$ and $v$ are the Higgs mass and vacuum expectation value
respectively\footnote{In \eq{sigma0Higgs} we have omitted scale factors relating 
to the $d$-dimensional coupling $\alpha_s$, which amounts to the choice 
$\overline{\mu} = m_h$.}. The form factor $F(\tau,\e)$ depends on the 
dimensionless variable
\beq
  \tau \, = \, \frac{s}{4 m_t^2} \, ,
\label{taudef}
\eeq
and it is given by~\cite{Dawson:1991zj}
\beq
  F (\tau, \e) \, = \, \frac{9}{4} \frac{1}{\tau^2} \left| 1 + \left(1 - \frac{1}{\tau} \right) 
  \arcsin^2 \left( \sqrt{\tau} \right) \right|^2 \, + \, {\cal O} (\e) \, ,
\label{formfactau}
\eeq 
with a normalisation chosen so that $F(\tau, \e) \to 1$ as $\tau
\to 0$. The cross section with kinematics shifted according to
\eq{sshift} can then be written as \beq \sigma^h_{\rm Born} ( z s ) \,
= \, \frac{\alpha_s^2}{\pi} \, \frac{z}{576 v^2} \, (1 + \e) \, F
\left(z \tau, \e \right) \delta \left(z - \frac{m_h^2}{s} \right) \, .
\label{sigma0Higgs2}
\eeq
Substituting this result into \eq{simpfact} and expanding in powers of $(1- z)$ 
and $\e$ one finds
\beq
  \frac{d \sigma^h_{\rm NLP}}{d z} \, = \, \frac{\alpha_s^3 C_A}{288 \pi^2 v^2} \, 
  F (z \tau, \e) \left( \frac{2 - {\cal D}_0 (z)}{\e} + 2 {\cal D}_1 (z) - {\cal D}_0 (z)
  - 4 \log(1 - z) + 2 \right) \, .
\label{dsigmadzres}
\eeq It is easy to check that \eq{dsigmadzres} agrees with the known
analytic NLO result of Ref.~\cite{Dawson:1991zj} in the $m_t \to
\infty$ limit. We note, however, that the result of \eq{dsigmadzres}
is much more informative: it includes the full dependence on the top
quark mass up to NLP order, and can thus be applied for arbitrary
$m_t$.  This is a remarkable simplification of the intricate result of
Ref.~\cite{Spira:1995rr} for the full $m_t$ dependence: after shifting
the kinematics of the leading order result, the resulting K-factor is
entirely independent of the top quark mass, which makes the formula
especially simple to apply in practical applications~\footnote{Indeed,
  we have checked that eq.~(\ref{dsigmadzres}) reproduces the K-factor
  reported in ref.~\cite{Harlander:2009mq}, which features a double
  expansion in threshold parameter and top mass.}.

It is interesting to examine the anatomy of the result in \eq{dsigmadzres} in slightly 
more detail. If one were to calculate the NLO cross section by starting manifestly 
in the large top mass limit ({\it i.e.} by using an effective field theory), the leading 
order graph would contain an effective point-like interaction coupling the two 
incoming gluons to a Higgs, as shown in Fig.~\ref{fig:singleHiggs}(b). At NLO, one 
can radiate the extra gluon from either of the incoming gluons, and one must also 
include the additional effective coupling shown in Fig.~\ref{fig:singleHiggs}(c), 
namely a point-like interaction between three gluons and a Higgs boson. If one 
resolves the top quark loop as in Fig.~\ref{fig:singleHiggs}(a), this extra interaction
corresponds to emissions from inside the top quark loop. In the NLP calculation, 
there is no need to include any additional diagrams to capture these contributions: 
they are generated precisely by the orbital angular momentum contributions in
\eq{AderivdiHiggs}: therefore, as the above analysis reveals, we can choose to 
associate these terms with a shift in the kinematics of the leading order result, up 
to corrections subleading in soft momentum. Seen from the point of view of the 
effective field theory at large $m_t$, it is highly non-trivial that such a shift captures 
the contribution of higher-order operators in the effective Lagrangian.


\section{Multiple Higgs boson production}
\label{sec:double}

\begin{figure}
\begin{center}
  \scalebox{0.8}{\includegraphics{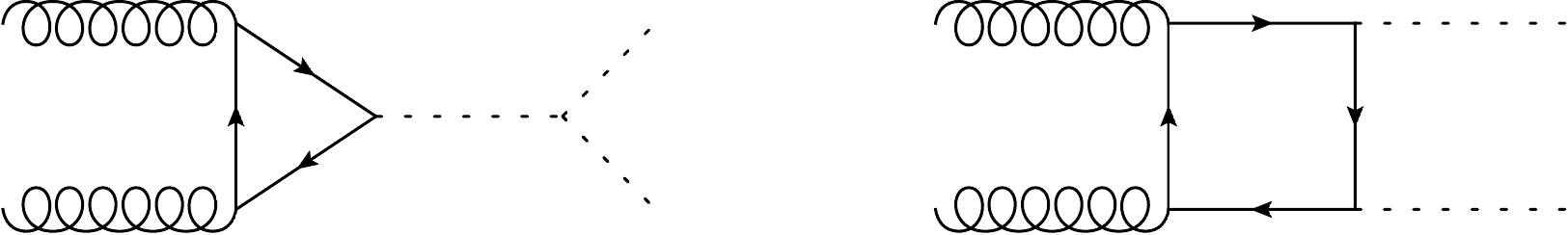}}
  \caption{Leading-order diagrams contributing to Higgs boson pair production.}
\label{fig:diHiggs}
\end{center}
\end{figure}

In the previous section we have tested our main result, given by \eq{simpfact} for 
gluon scattering, by reproducing known results in the cross section for single Higgs 
boson production via gluon fusion. We now consider the case of Higgs boson 
pair production, a process of ongoing interest at the LHC, due to its potential 
role in extracting the Higgs boson self-coupling. Analytic results for this process 
are known up to NNLO in the large top mass limit~\cite{Dawson:1998py,
deFlorian:2013jea,deFlorian:2016uhr}, but only at leading order with full top mass 
dependence~\cite{Glover:1987nx,Plehn:1996wb}. Further studies have looked at 
systematically improving the effective field theory results by including leading-power 
threshold effects~\cite{Shao:2013bz}, or contributions suppressed by powers of the 
top mass~\cite{Grigo:2013rya}. Recently, numerical results at NLO accuracy with 
full top mass dependence have become available~\cite{Borowka:2016ypz} (see
also~\cite{Frederix:2014hta}). This, however, does not preclude the desire for 
analytic results, which can serve to improve the efficiency of numerical computations, 
whilst also providing clues regarding higher-order structures in perturbation theory. 
This is especially true in Higgs boson pair production, given that the large top mass 
limit is not a good approximation, unlike the case of single Higgs production. The 
leading order diagrams for Higgs pair production are shown in Fig.~\ref{fig:diHiggs}, 
and the leading order amplitude may be written as
\beq
  {\cal A}_{\mu \nu} \, = \, \frac{\alpha_s}{2 \pi v^2} \Big[
  \left( C_\triangle F_\triangle + C_\Box F_\Box \right) \, T_{0, \, \mu \nu} + 
  C_\Box G_\Box \, T_{2, \, \mu \nu} \Big] \, ,
\label{ALO}
\eeq
where (in the Standard Model)
\beq
  C_\triangle\, = \, \frac{3 m_h^2}{s - m_h^2} \, , \qquad C_\Box \, = \, 1 \, ,
\label{Ctrisqdef}
\eeq
and $F_\triangle$, $F_\Box$, $G_\Box$ are form factors arising from the triangle 
and box graphs, as indicated by the subscripts. They depend on the Higgs boson 
and top masses, as well as the partonic centre of mass energy $s$ and the 
other Mandelstam invariants
\beqa
  t & = & - \frac12 \left[ s - 2 m_h^2 - 
  s\sqrt{1-\frac{4m_h^2}{s}} \cos \theta \right] \, , \nonumber \\
  u & = & - \frac12 \left[ s - 2 m_h^2 + 
  s\sqrt{1-\frac{4m_h^2}{s}} \cos \theta \right] \, .
\label{mandies}
\eeqa
The basis tensors $T_{s, \, \mu \nu}$, with $s = 0,2$, 
in \eq{ALO} are associated with the exchange of spin 0 and spin 2 in the $s$ 
channel, respectively. Denoting the gluon momenta by $p_1$ and $p_2$ and the
Higgs boson momenta by $p_3$ and $p_4$, their explicit forms are
\beqa
  T_0^{\mu \nu} & = & p_1 \cdot p_2 \, \eta^{\mu \nu} - p_1^\nu \, p_2^\mu \, ,
  \label{A1A2def} \\
  T_2^{\mu \nu} & = & p_1 \cdot p_2 \, \eta_{\mu \nu} + \frac{1}{p_T^2}
  \Big[ m_h^2 \, \, p_1^\nu \, p_2^\mu - 2 p_2 \cdot p_3 \, p_1^\nu \, p_3^\mu
  - 2 p_1 \cdot p_3 \, p_2^\mu \, p_3^\nu + 2 p_1 \cdot p_2 \, 
  p_3^\mu \, p_3^\nu \Big] \, , \nonumber 
\eeqa
where
\beq
  p_T^2 \, = \, \frac{2 (p_1 \cdot p_3)(p_2 \cdot p_3)}{p_1 \cdot p_2} - m_h^2 \, .
\label{pT2def}
\eeq
With these notations, the leading-order distribution in the Mandelstam invariant 
$t$ can be written as~\cite{Dawson:1998py}
\beq
  \frac{d \hat{\sigma}_{\rm Born}^{hh}}{d t} \, = \, \frac{\alpha_s^2}{8 \pi^3} \, 
  \frac{1}{512 \, v^4} \left[ \big| C_\triangle F_\triangle + C_\Box F_\Box \big|^2
  + \big| C_\Box G_\Box \big|^2 \right] \, .
\label{sigma0diHiggs}
\eeq
This expression simplifies considerably in the large top mass limit, where
\beq
  F_\triangle \, \to \, \frac{2}{3} \, , \qquad 
  F_\Box \, \to \, - \frac{2}{3} \, , \qquad
  G_\Box \, \to \,  0 \, ,
\label{FGlims}
\eeq
so that \eq{sigma0diHiggs} becomes
\beq
  \frac{d \hat{\sigma}^{hh}_{\rm Born}}{d t} \, = \, 
  \frac{\alpha_s^2}{16 \pi^3} \, \frac{1}{576 \, v^4}
  \left( \frac{4m_h^2 - s}{s - m_h^2} \right)^2 \, .
\label{sigma0diHiggs2}
\eeq We observe that, in the large top mass limit, the leading-order
cross section vanishes at threshold, as $s \to 4 m_h^2$, due to the
cancellation between the box and triangle contributions. This property
is one of the reasons that make the large top mass limit a poor
approximation in Higgs boson pair production, necessitating the
calculation of higher order corrections with full $m_t$ dependence.
It also causes problems when trying to define a K-factor as a function
of the variable $z$. Ordinarily, one would divide the NLO cross
section by the LO one, however this becomes ill-defined in the
threshold region $z \to 1$. In Ref.~\cite{Dawson:1998py} this problem
is circumvented by dividing by the LO cross section with kinematics
shifted according to \eq{sshift2}. The resulting K-factor thus
matches precisely the quantity defined in \eq{simpfact}.

With this convention, the NLO cross section for Higgs boson pair production, 
up to NLP accuracy, can be written as 
\beq
  z \, \frac{d \sigma_{\rm NLP}^{hh}}{d z} \, = \, \frac{\alpha_s}{3\pi} \, C_A \, 
  \left( \frac{\overline{\mu}^2}{s} \right)^\e
  \left[ \frac{12 - 6 {\cal D}_0 (z)}{\e} + 12 {\cal D}_1 (z) - 24 \log (1 - z) \right]
  \sigma^{hh}_{\rm Born} \left( z s \right) \, ,
\label{diHiggsNLP}
\eeq
where we have extracted an explicit factor of $z$ on the left-hand side, to 
match the conventions adopted in Ref.~\cite{Dawson:1998py}. In the large top 
mass limit, \eq{diHiggsNLP} reproduces the results of Ref.~\cite{Dawson:1998py}.
As in the case of single Higgs production, however, the result is much more 
powerful, in that it applies to the full top mass dependence. \eq{diHiggsNLP} 
thus provides an explicit analytic form of the cross section, at NLO in perturbation 
theory, and up to NLP in the threshold variable $(1 - z)$. This extends the 
results of Ref.~\cite{Shao:2013bz}, which considered supplementing the 
fixed-order cross section with threshold effects at leading power only. 
Ref.~\cite{Grigo:2013rya}, on the other hand, studied the systematic 
improvement of the large top mass limit, by including corrections expressed 
as a power series in $m_h^2/m_t^2$. The authors found that the 
convergence of this expansion could be dramatically improved by factorising 
the leading order cross-section with exact top mass dependence. The results 
of this paper explain why this is so: indeed, we find that, to first subleading 
order in the threshold expansion, the NLO cross section can be completely
expressed in terms of the leading-order cross section with shifted kinematics, 
and with full top mass dependence.

Returning to the large top mass limit, we can go further and consider triple Higgs 
production, a process for which analytic NLO corrections were presented recently 
in Ref.~\cite{deFlorian:2016sit}. If we denote the invariant mass of the triple 
Higgs system by $M_{3 h}$, and we set the $\overline{\rm MS}$ scale according 
to $\mu^2 = M_{3 h}^2$, the NLO cross section in the gluon fusion channel can
be written as~\cite{deFlorian:2016sit}
\beq
  M_{3 h}^2 \, \, \frac{d \sigma^{hhh} }{d M_{3 h}^2} \, = \, \frac{\alpha_s}{2 \pi} \,
  \sigma^{hhh}_{\rm Born} \left( z s \right) \, \eta (z) \, ,
\label{tripleNLO}
\eeq
with
\beqa
  \eta (z) & = & 24 {\cal D}_1 (z) - 24 z \left(- z + z^2 + 2 \right) \log(1 - z) 
  \nonumber \\ 
  && \quad - \, \frac{12 (z^2 + 1 - z)^2}{1 - z} \log(z) - 11 (1 - z)^3 + C^{(\delta)}_{3 h} 
  \delta(1 - z) \, ,
\label{etaggdef}
\eeqa
where $C^{(\delta)}_{3 h}$ can be read off from Ref.~\cite{deFlorian:2016sit}, 
and does not affect our arguments. Expanding to NLP in $(1 - z)$, we may 
rewrite \eq{tripleNLO} as
\beq
  z \, \frac{d \sigma^{hhh}}{d z} \, = \, \frac{\alpha_s}{4 \pi} \, C_A \, 
  \Big[ 16 {\cal D}_1 (z) - 32 \log(1 - z) + 8 + {\cal O} (1 - z) \Big] \,
  \sigma^{hhh}_{\rm Born} \left( z s \right) \, .
\label{tripleNLO2}
\eeq
This result is indeed reproduced from \eq{simpfact}: implementing the scale choice 
as in \eq{scales},  one can rewrite \eq{simpfact} in the present  case as
\beq
  z \, \frac{d \sigma^{hhh}}{d z}  \, = \, \frac{\alpha_s}{4 \pi} \, C_A \, \left[ 
  \frac{16 - 8 {\cal D}_0 (z)}{\e} + 16 {\cal D}_1 (z) - 32 \log(1 - z) + 8 \right] \, 
  \sigma^{hhh}_{\rm Born} \left( z s \right) \, ;
\label{Kfachhh}
\eeq
extracting the finite part in the $\overline{\rm MS}$ scheme, one finds precise 
agreement with \eq{tripleNLO2}.


\section{Vector boson pair production}
\label{sec:diphoton}

In the preceding two sections, we have illustrated the application of our general 
expression for the NLO K-factor in the gluon channel. In order to verify and illustrate
the quark result, \eq{simpfact2}, in a non-trivial case, it is instructive to see how 
known matrix elements in vector boson pair production can be reproduced. This
calculation is also an important illustration of the fact that our prediction, while
based on power counting in the threshold variable $z$, applies to the fully differential
squared amplitude, and not only to the integrated cross section. As an example, 
we consider di-boson production,
\beq
  q (p_1) + \bar{q} (p_2) \, \rightarrow \, V(p_3) + V(p_4) \, ,
\label{Vpair}
\eeq
where $V$ is an electroweak gauge boson. Let us start with di-photon production, 
the amplitudes for which can be found in Ref.~\cite{Gastmans:1990xh} up to NLO, 
in four spacetime dimensions. The squared LO matrix element for this process, 
summed and averaged over colours and spins, is given by
\beq
  \overline{ \left| A^{\gamma \gamma}_{\rm Born} \right|^2} \, = \, 
  \frac{2 e_q^4}{N_c} \, \frac{t^2 + u^2}{t u} \, = \,
  \frac{4 e_q^4}{N_c} \, \frac{1 + \cos^2 \theta}{1 - \cos^2 \theta} \, ,
\label{ALOdiphoton}
\eeq
where $e_q$ is the electromagnetic charge of the quark. One may now consider 
the radiation of an additional gluon, leading to the process
\beq
  q (p_1) + \bar{q} (p_2) \, \rightarrow \, V (p_3) + V (p_4) + g (k) \, .
\label{Vpair2}
\eeq
Computing the di-photon cross section at NLO, one must include squared Born 
matrix element for the process in \eq{Vpair2}, again summed and averaged over 
colours and spins. It is given by~\cite{Gastmans:1990xh}
\beq
  \overline{ \left| A_{\rm NLO}^{\gamma \gamma g} \right|^2} \, = \, 
  \frac{e_q^4}{N_c} \, g_s^2 C_F \, \frac{s \, \sum_{i} ( p_1 \cdot k_i ) 
  ( p_2 \cdot k_i ) \left[ (p_1 \cdot k_i)^2 + (p_2 \cdot k_i)^2 \right]}{\prod_{i}
  ( p_1 \cdot k_i ) ( p_2 \cdot k_i ) } \, , 
  \qquad k_i \in \{ p_3, p_4, k \} \, .
\label{ANLOgam}
\eeq
Our aim is to show that, up to NLP accuracy, this matrix element can be obtained 
by shifting kinematics in the LO squared amplitude, as dictated by \eq{ANLP3}. 
To this end, one may rescale the gluon momentum as $k \to \lambda k$ in 
\eq{ANLOgam}, and expand to next-to-leading power in $\lambda$, before 
setting $\lambda \to 1$. Next, one can parametrise the momenta in the centre 
of mass frame of the $VV$ system, as (see for example Ref.~\cite{Frixione:1993yp})
\beqa
  p_1 & = & p_1^0 \, (1, 0, 0, 1) \, ,  \quad 
  p_2 \,\, = \,\, p_2^0 \, (1, 0, \sin \psi, \cos \psi) \, , \quad
  k \,\, = \,\, k^0 \, (1, 0, \sin \psi', \cos \psi') \nonumber \\
  p_3 & = & \frac{\sqrt{s_2}}{2} \, \big(1, \beta_z \sin \theta_2 \sin \theta_1,
  \beta_z \cos \theta_2 \sin \theta_1, \beta_z \cos \theta_1 \big) \, , \nonumber \\
  p_4 & = & \frac{\sqrt{s_2}}{2} \, \big(1, - \beta_z \sin \theta_2 \sin \theta_1, 
  - \beta_z \cos \theta_2 \sin \theta_1, - \beta_z \cos \theta_1 \big) \, ,
\label{momparamgam}
\eeqa
where
\beqa
  && p_1^0 \, = \, \frac{s + t_k}{2 \sqrt{s_2}} \, , \quad 
        p_2^0 \, = \, \frac{s + u_k}{2 \sqrt{s_2}} \, , \quad 
        k^0 \, = \, - \frac{t_k + u_k}{2 \sqrt{s_2}} \, , \nonumber \\
  && \cos \psi \, = \, 1 - \frac{s}{2 p_1^0 p_2^0} \, , \quad 
        \cos \psi' \, = \, 1 + \frac{t_k}{2 p_1^0 k^0} \, , \quad 
        \beta_z \, = \, \sqrt{1 - \frac{4 m_V^2}{z s}} \, ,
\label{momparamgam2}
\eeqa
and we have introduced the invariants
\beq
  t_k \, = \, (p_1 - k)^2 \, , \qquad  u_k \, = \, (p_2 - k)^2 \, , \qquad 
  s_2 \, = \, s + t_k + u_k \, .
\label{mandies2}
\eeq
Note that in these notations the threshold variable is $\xi = 1 - z = 1 - s_2/s$. With 
these definitions, we can now expand to first subleading power in the gluon energy, 
and set $m_V = 0$ for the di-photon case. The result is
\beq
  \overline{\left| A_{\rm NLP}^{\gamma \gamma g} \right|^2} \, = \, g_s^2 C_F \,
  \frac{s}{p_1 \cdot k \, p_2 \cdot k} \, \frac{4 e_q^4}{N_c} \, \bigg[ 
  \frac{1 + \cos^2 \theta_1}{1 - \cos^2 \theta_1} - 
  \frac{8 \sin \theta_1 \cos \theta_1 \cos \theta_2}{ \left(1 - \cos^2 \theta_1 \right)^2}
  \frac{\sqrt{p_1 \cdot k \, p_2 \cdot k}}{s} \bigg] \, ,
\label{ANLPgamB}
\eeq
By performing the same procedure, one may easily show that the LO amplitude of 
\eq{ALOdiphoton}, evaluated with the kinematic shifts defined in \eq{deltapdef}, yields
\beq
  \overline{ \left| A_{\rm Born}^{\gamma \gamma} \left( p_1 + \delta p_1,
  p_2 + \delta p_2 \right) \right|^2} \, = \, \frac{4 e_q^4}{N_c} \, 
  \bigg[ \frac{1 + \cos^2 \theta_1}{1 - \cos^2 \theta_1} - 
  \frac{8 \sin \theta_1 \cos \theta_1 \cos \theta_2}{(1 - \cos^2 \theta_1)^2}
  \frac{\sqrt{p_1 \cdot k \, p_2 \cdot k}}{s} \bigg] \, .
\label{ALOgamshiftB}
\eeq 
We therefore see that \eq{ANLPgamB} explicitly confirms the
expectations raised by \eq{ANLP3}.

A similar exercise may be carried out for $W^+W^-$ production: the NLO squared 
amplitudes for this process (again in four spacetime dimensions) may be found 
in Ref.~\cite{Frixione:1993yp}. The Born-level squared matrix element, summed 
and averaged over colours and spins, is given by
\beq
  \overline{ \left| A_{\rm Born}^{W W} \right|^2} \, = \, 
  \frac{1}{4 N_c} \left[ c^{tt}_i \, F_i^{(0)} (s, t)
  - c^{ts}_i (s) \, J_i^{(0)} (s, t)
  + c^{ss}_i (s) \, K_i^{(0)} (s, t) \right] \, ,
\label{LOWW}
\eeq
where $c^{tt}_i$, $c^{ss}_i$ and $c^{ts}_i$ are coefficients associated with 
$t$-channel, $s$-channel and interference graphs, respectively, for a quark of 
flavour $i$, and can be found in Ref.~\cite{Frixione:1993yp}. The remaining 
functions of Mandelstam invariants are given by
\beqa
  F_i^{(0)} (s, t) & = & 16 \left( \frac{u t}{m_W^4} - 1 \right)
  \left( \frac{1}{4} + \frac{m_W^4}{t^2} \right)
  + 16 \, \frac{s}{m_W^2} \, , 
  \nonumber \\
  J_i^{(0)} (s, t) & = & 16 \left( \frac{u t}{m_W^4} - 1 \right)
  \left( \frac{s}{4} - \frac{m_W^2}{2} - \frac{m_W^4}{t} \right)
  + 16 s \left( \frac{s}{m_W^2} - 2 + 2 \, \frac{m_W^2}{t} \right) \, ,
  \nonumber \\
  K^{(0)}_i (s, t) & = & 8 \left( \frac{u t}{m_W^4} - 1 \right)
  \left( \frac{s^2}{4} - s m_W^2 + 3 m_W^4 \right)
  + 8 s^2 \left( \frac{s}{m_W^2} - 4 \right) \, .
\label{WWfuncs}
\eeqa
Similarly, the NLO squared matrix element including the radiation of a gluon, for 
the $q \bar{q}$ initial state, also summed and averaged over colours and spins, 
is given by
\beq
  \overline{ \left| A_{\rm NLO}^{WWg} \right|^2} \, = \,
  \frac{4 \pi \alpha_s C_F}{N_c} \, \frac{s}{t_k u_k} \,  
  \Big[ c_i^{tt} \, \widehat{X}_i - c_i^{ts} (zs) \, \widehat{Y}_i + 
  c_i^{ss} (zs) \, \widehat{Z}_i \Big] \, ,
\label{WWNLO}
\eeq
where the functions $\widehat{X}_i$, $\widehat{Y}_i$ and $\widehat{Z}_i$ have 
been obtained by rescaling the corresponding functions  $X_i$, $Y_i$ and $Z_i$,
given in Appendix D of Ref.~\cite{Frixione:1993yp}, extracting the singular prefactor
$- 4 s/(t_k u_k)$. Parametrising momenta as in \eq{momparamgam}, one may 
expand each function up to NLP in $\xi = (1 - z)$, using the same procedure as 
outlined above for the di-photon matrix element. Introducing the notation $\rho = 
m_W^2/s$, the results can be written as
\beqa
  \widehat{X}_i \Big|_{\rm NLP} & = & \frac{32}{ (4 \rho - 1) \big[ 1 - 2 \rho 
  + (4 \rho - 1) \cos \theta_1 \big]^3} \, \bigg\{ \cos^2 \theta_1 \left(32 \rho^3 
  - 32 \rho^2 + 10 \rho - 1 \right) 
  \nonumber \\
  & & + \, \cos \theta_1 \left( 96 \rho^4 - 112 \rho^3
  + 70 \rho^2 - 20 \rho + 2 \right) - 16 \rho^4 + 44 \rho^3 - 34 \rho^2
  \nonumber\\
  & & + 10 \rho - 1 + \xi \, \Big[ \cos^2 \theta_1 \left( 128 \rho^4 - 96 \rho^3 
  + 24 \rho^2 - 2 \rho \right)
  + \cos \theta_1 \left( 32 \rho^3 \right.
  \nonumber \\
  & & \left. - 16 \rho^2 + 2 \rho \right) + 16 \rho^3 - 4 \rho^2 \Big] \bigg\}
  \, - \, \frac{\cos^2 \theta_1}{(4 \rho - 1)}
  \bigg[ 192 \rho - 112 + \frac{20}{\rho} - \frac{1}{\rho^2} 
  \nonumber \\
  & & + \, \xi \, \left( 128 \rho - 96 + \frac{24}{\rho} - \frac{2}{\rho^2} \right)
  \bigg] + \frac{1}{\rho^2} + \frac{12}{\rho} - 16
  + \, \xi \, \left( \frac{2}{\rho^2} + \frac{12}{\rho} \right)
  \label{XNLP} \\
  & & + \frac{\sin \psi \cos \theta_2 \sin \theta_1}{\big[1 - 2 \rho + (4 \rho - 1) 
  \cos \theta_1 \big]^3} \, \Bigg[ \cos^4 \theta_1 \, \bigg( 1024 \rho^3 - 1280 \rho^2
  + 640 \rho - 160 
  \nonumber \\
  & & + \frac{20}{\rho} - \frac{1}{\rho^2} \bigg) 
  + \cos^3 \theta_1 \left( - 1536 \rho^3 + 2304 \rho^2 - 1344 \rho + 384 
  - \frac{54}{\rho} + \frac{3}{\rho^2} \right) 
  \nonumber \\
  & & + \cos^2 \theta_1 \left( 768 \rho^3 - 1344 \rho^2 + 912 \rho - 300 
  + \frac{48}{\rho} - \frac{3}{\rho^2} \right) \nonumber \\
  & & + \cos \theta_1 \left( - 640 \rho^3 + 768 \rho^2 - 360 \rho + 92 
  - \frac{14}{\rho} + \frac{1}{\rho^2} \right) - 256 \rho^2 + 128 \rho - 16 \Bigg] \, ,
  \nonumber
\eeqa
\beqa
  \widehat{Y}_i \Big|_{\rm NLP} & = & \frac{32 s \rho}{\big[ 1 - 2 \rho + (4 \rho - 1) 
  \cos \theta_1 \big]^2} \bigg[ \cos \theta_1 (\rho + 2) + 2 \rho^2 
  + 3 \rho - 2
  \nonumber\\
  & & + \, \xi \rho \, \Big( \cos \theta_1 (4 \rho - 1) + 5 \Big) \bigg]
  + s \, \bigg\{ \cos^2\theta_1 \left( 96 \rho - 80 + \frac{18}{\rho} 
  - \frac{1}{\rho^2} \right)
  \nonumber\\
  & & + \, 8 \cos \theta_1 \left( - 8 \rho + 1 \right) - 16 \rho - 16 + \frac{10}{\rho} 
  + \frac{1}{\rho^2} + \xi \, \bigg[ \cos^2 \theta_1 \left( 64 \rho - 80 + \frac{28}{\rho} 
  - \frac{3}{\rho^2} \right)
  \nonumber\\
  & & + 8 \cos \theta_1 \left( - 4 \rho + 1 \right) - 16 + \frac{20}{\rho} 
  + \frac{3}{\rho^2} \bigg] \bigg\} 
  \nonumber \\ 
  & & + \, \frac{s \sin \psi \cos \theta_2 \sin \theta_1}{ \big[ \left(4 \rho - 1 \right) 
  \cos \theta_1 + 1 - 2 \rho \big]^2} \,\Bigg[ \cos^3 \theta_1 \left( - 512 \rho^3
  + 768 \rho^2 - 448 \rho + 128 - \frac{18}{\rho} + \frac{1}{\rho^2} \right)
  \nonumber \\
  & & + \, \cos^2\theta_1 \left( 768 \rho^3 - 1088 \rho^2  
  + 656 \rho - 204 + \frac{32}{\rho} - \frac{2}{\rho^2} \right)
  \nonumber\\
  & & + \, \cos\theta_1 \, \left( - 384 \rho^3 + 512 \rho^2 - 280 \rho + 84 - \frac{14}{\rho} 
  + \frac{1}{\rho^2} \right) - 192 \rho^2 + 64 \rho - 4 \Bigg] \, ,
\label{YNLP}
\eeqa
\beqa
  \widehat{Z}_i \Big|_{\rm NLP} & = & s^2 \, \Bigg\{ \cos^2 \theta_1 \left(
  - 288 \rho^2 + 192 \rho - 62 + \frac{10}{\rho} - \frac{1}{2 \rho^2} \right)
  - 24 \rho - 18 + \frac{4}{\rho} + \frac{1}{2 \rho^2} 
  \nonumber \\
  & & + \, \xi \, \bigg[ \cos^2 \theta_1 \left( -192 \rho^2 + 192 \rho
  - 92 + \frac{22}{\rho} - \frac{2}{\rho^2} \right)  
  - 24 \rho - 36 + \frac{12}{\rho} + \frac{2}{\rho^2} \bigg] \Bigg\}
  \nonumber \\
  & & + \, s^2 \, \sin \psi \cos \theta_2 \sin \theta_1 \cos \theta_1
  \left( 96 \rho^2 - 80 \rho + 30 - \frac{6}{\rho} + \frac{1}{2 \rho^2} \right) \, .
\label{ZNLP}
\eeqa
We have explicitly checked that the same results are obtained from \eq{ANLP3}, 
where for the right-hand side one must use the LO squared amplitude given in
\eq{LOWW}, with momenta shifted according to \eq{deltapdef}. This is a highly 
non-trivial cross-check: comparing Eqs.~(\ref{XNLP} - \ref{ZNLP}) with the di-photon 
case given in \eq{ANLPgamB}, one sees that the $WW$ case involves a much more
complicated dependence on the opening angle $\theta_1$, and the
partonic centre of mass energy $s$. A similar analysis could be
carried out for $ZZ$ production~\cite{Mele:1990bq}, and also to
provide analytic information for triple vector boson production,
numerical results for which have been presented in
refs.~\cite{Lazopoulos:2007ix,Campanario:2008yg}.


\section{Conclusion}
\label{sec:conclude}

In this paper, we have considered the hadro-production of an arbitrary heavy 
colourless system, in both the gluon-fusion and quark-antiquark-annihilation
channels, near partonic threshold for the production of the selected final state.
Our starting point is an all-order factorisation formula for the relevant scattering 
amplitudes, introduced in Refs.~\cite{Bonocore:2015esa,Bonocore:2016awd}, 
given here in \eq{NEfactor_nonabel}, and valid to next-to-leading power 
in the threshold expansion. Specialising this formula to NLO in the strong 
coupling, we have observed how the general expression simplifies, and takes 
the form of a next-to-soft theorem, as derived for example in~\cite{Casali:2014xpa,
White:2011yy,Cachazo:2014fwa}. This simple expression, in turn, leads to
a universal form for the transition probability, completely differential in the final 
state variables, and proportional to the Born-level transition probability, computed  
with a specific shift for the initial parton momenta. The result is the same for
quarks and gluons (up to a trivial substitution of color factors), and is reported
in Eqs.~(\ref{ANLPshift}, \ref{ANLP3}). When the transition probability is integrated
over final state variables, one finds that the inclusive cross section for the
selected process can also be written in a simple and universal factorised form,
given here in Eqs.~(\ref{simpfact}, \ref{simpfact2}). More precisely, at NLO, and 
up to next-to-leading power in the threshold expansion, the cross-sections can be 
written as a universal $K$-factor, multiplying the leading order cross-section with 
a shifted partonic centre-of-mass energy. All these results apply regardless of 
whether the leading order process is tree-level or loop-induced, and the resulting 
K-factors are independent of hard scales such as heavy quark masses.

We have checked our results by reproducing known expressions for the production 
of up to three Higgs bosons at NLO, in the large top mass limit. Away from this limit, 
our formula provides new analytic information in the case of Higgs pair production at 
NLO, where only numerical results are presently known. Furthermore, we explain the
observation, made previously in Ref.~\cite{Grigo:2013rya}, that the convergence of 
the large top mass expansion can be improved in this process by factoring out the 
LO cross-section with exact top mass dependence. In the quark channel, we have 
shown how our formula is consistent with previous results for the production of 
photon and $W$ boson pairs, again at the level of differential distributions.

The results we have presented show that, to NLP accuracy, differential and inclusive
NLO cross sections for colour-singlet final states are dramatically simpler than
exact results, and we expect that they will be very easy to implement in numerical 
codes, providing checks of existing calculations, and improved approximations
for differential distributions when complete results are not available, as is the case
for loop induced processes with multi-particle electroweak final states. A detailed
phenomenological analysis, including a study of the accuracy of the NLP approximation
in different processes and kinematic domains, has been left to future work.

We emphasise that the simple universal expressions that we find at NLO
can be systematically improved upon by relying on
\eq{NEfactor_nonabel}: in particular, a NLO calculation of the
radiative soft function and of the radiative jet functions for quarks
and gluons, which is under way, will lead to NLP approximations for
cross sections of the type studied in this paper at NNLO level,
exploring uncharted territory, in particular for processes which are
loop induced and have so far been studied predominantly in the context
of effective field theory approximations.  We also note that the
universal expressions we have derived for inclusive cross sections are
also applicable to colourless final states arising beyond the Standard
Model, in which case our results can provide a controlled
approximation to estimate the size of higher-order corrections in
selected models without the need to perform calculations that would be
expensive for loop-induced processes.


\section*{Acknowledgments}

We thank Fabrizio Caola, Jort Sinninghe-Damst\'{e}, Falko Dulat,
Bernhard Mistlberger and Adrian Signer for discussions. We also thank
Michael Spira for kindly providing us with personal notes relevant to
our results in Section 3.  This research was supported by the Research
Executive Agency (REA) of the European Union through the contract
PITN-GA-2012-316704 (HIGGSTOOLS).  EL was supported by the Netherlands
Foundation for Fundamental Research of Matter (FOM) programme 156,
``Higgs as Probe and Portal'', and by the National Organisation for
Scientific Research (NWO). CDW is supported by the UK Science and
Technology Facilities Council (STFC), and thanks ETH, Zurich and the
Higgs Centre for Theoretical Physics, University of Edinburgh, for
hospitality.  LV is supported by the People Programme (Marie Curie
Actions) of the European Union Horizon 2020 Framework
H2020-MSCA-IF-2014, under REA grant N.~656463 (Soft Gluons).


\appendix


\bibliography{refs.bib}


\end{document}